\documentclass[english]{article}%
\usepackage[T1]{fontenc}
\usepackage[latin1]{inputenc}
\usepackage{amsmath}
\usepackage{amsmath}
\usepackage{graphicx}
\usepackage{babel}%
\usepackage{amsfonts}%
\usepackage{amssymb}
\renewcommand{\mathbf}[1]{\mbox{\boldmath$#1$}}

\makeatletter
\newtheorem{theorem}{Theorem}
\newtheorem{acknowledgement}[theorem]{Acknowledgement}
\makeatother

\begin{document}

\title{{\Large General quadratic gauge theory. Constraint structure, symmetries, and
physical functions. }}
\author{D.M. Gitman\thanks{Institute of Physics, University of Sao Paulo, Brazil;
e-mail: gitman@fma.if.usp.br} \ and I.V. Tyutin\thanks{Lebedev Physics
Institute, Moscow, Russia; e-mail: tyutin@lpi.ru}}
\maketitle

\begin{abstract}
How can we relate the constraint structure and constraint dynamics of the
general gauge theory in the Hamiltonian formulation with specific features of
the theory in the Lagrangian formulation, especially relate the constraint
structure with the gauge transformation structure of the Lagrangian action?
How can we construct the general expression for the gauge charge if the
constraint structure in the Hamiltonian formulation is known? Whether can we
identify the physical functions defined as commuting with first-class
constraints in the Hamiltonian formulation and the physical functions defined
as gauge invariant functions in the Lagrangian formulation? The aim of the
present article is to consider the general quadratic gauge theory and to
answer the above questions for such a theory in terms of strict assertions. To
fulfill such a program, we demonstrate the existence of the so-called
superspecial phase-space variables in terms of which the quadratic Hamiltonian
action takes a simple canonical form. On the basis of such a representation,
we analyze a functional arbitrariness in the solutions of the equations of
motion of the quadratic gauge theory and derive the general structure of
symmetries by analyzing a symmetry equation. We then use these results to
identify the two definitions of physical functions and thus prove the Dirac conjecture.

\end{abstract}

\section{Introduction}

The most of contemporary particle-physics theories are formulated as gauge
theories. It is well known that within the Hamiltonian formulation, gauge
theories are theories with constraints (in particular, with first-class
constraints (FCC)). This is the main reason for a long intensive study formal
theory of constrained systems. The theory of constrained systems has began
with pioneer works by Bergman and Dirac \cite{Berg,Dirac} and was then
developed and presented in some review books
\cite{Sudar74,HanReT76,Sunde82,GitTy90,HenTe92} still attracts big attention
of researchers. Were relatively simple were the first steps of the theory in
formulating dynamics of constrained systems in the phase space, elaborating
the procedure of finding all the constraints (the Dirac procedure), and
reorganizing the constraints to the FCC and second-class constraints (SCC).
From the very beginning, it became clear that the presence of FCC among the
complete set of constraints in the Hamiltonian formulation is a direct
indication that the theory is a gauge one, i.e., its Lagrangian action is
invariant under gauge transformations that in general case, are continuos
transformations parametrized by arbitrary functions of time (of space-time
coordinates in the case of field theory). It was demonstrated that the number
of independent gauge parameters is equal to the number $\mu_{1}$ of primary
FCC and the total number of unphysical variables is equal to the number $\mu$
of all FCC, in spite of the fact that the equations of motion contain only
$\mu_{1}$ arbitrary functions of time (undetermined Lagrange multipliers to
the primary FCC), see \cite{GitTy90} and references therein. At the same time,
we proved that for a class of theories for which the constraint structure of
the whole theory and its quadratic approximation is the same and for which the
constraint structure does not change from point to point in the phase-space
(we call such theories perturbative ones), physical functions in the
Hamiltonian formulation have to commute with FCC. In a sense, this statement
can be identified with the so-called Dirac conjecture. All known until now
models where the Dirac conjecture does not hold are nonperturbative in the
above sense. After this preliminary progress in the theory of constrained
systems, it became clear that a natural and very important continuation of the
study is to try to relate the constraint structure and constraint dynamics of
a gauge theory in the Hamiltonian formulation with specific features of the
theory in the Lagrangian formulation, especially to relate the constraint
structure with the gauge transformation structure of the Lagrangian action.
One of the key problem here is the following: how to construct a general
expression for the gauge charge if the constraint structure in the Hamiltonian
formulation is known? Another principle question, closely related to the
latter one, is: whether can we identify the physical functions defined as
commuting with FCC in the Hamiltonian formulation and the physical functions
defined as gauge invariant functions in the Lagrangian formulation? Many
efforts were made in attempting to answer these questions, see for example,
\cite{Castell}. All previous considerations contain some restrictive
assumptions about the theory structure (in particular, about the constraint
structure), such that strictly proved answers to all the above questions are
still unknown for a general gauge theory (even belonging to the
above-mentioned perturbative class).

The aim of the present work is to consider a general quadratic gauge theory
and to answer the above questions for such a theory in terms of strict
assertions. The motivation is that for the majority of the perturbative gauge
theories, their behavior is in essence determined by the quadratic part of the
action, and the nonquadratic part is ''small'', in a sense. The constraint and
gauge structure of the complete theory and its quadratic approximation is the
same. Constraints of the complete theory differ from linear constraints of the
quadratic theory by ''small'' nonlinear terms, such that the number of
first-class and second-class constraints remains unchanged. The gauge
transformations of the complete theory and of its quadratic approximation have
the same number of gauge parameters. The majority of the properties of the
complete gauge theory and of its quadratic approximation are the same.
However, as was already mentioned above, the consideration of a general gauge
theory is sometimes a formidable task. At the same time, the simplifications
due to the quadratic approximation allow present the strict derivations and
illustrations of relations between the Hamiltonian and Lagrangian structures
of gauge theories. In particular, we establish the relation between the
constraint structure of the theory and the structure of its gauge
transformations, we represent the gauge charge as a decomposition in
constraints, we prove the Dirac conjecture and identify the physical functions
in Hamiltonian and Lagrangian formulations. To fulfill such a program, we
demonstrate the existence of the so-called superspecial phase-space variables
(sec. 2), in terms of which the quadratic Hamiltonian action takes a simple
canonical form. On the basis of such a representation, we analyze a functional
arbitrariness in the solutions of the equations of motion of the quadratic
gauge theory (sec. 3), and derive a general structure of symmetries by
analyzing a symmetry equation (sec. 4). In sec. 5, we use these results to
identify the two definitions of physical functions and thus prove the Dirac conjecture.

\section{Superspecial phase-space variables}

In this Section, we demonstrate the existence of the so-called superspecial
phase-space variables for which the total Hamiltonian of a general quadratic
gauge theory takes a simple canonical form.

First, we recall \cite{GitTy90} that there exists a canonical transformation
from the initial phase-space variables $\eta=\left(  q,p\right)  $ to the
special phase-space variables $\vartheta=\left(  \omega,Q,\Omega\right)  $
with the following properties. The constraint surface is described by the
equations $\Omega=0.$ The variables $\Omega$ are divided into two groups:
$\Omega=(\mathcal{P},U)$, where $U$ are all the SCC and $\mathcal{P}$ are all
the FCC. At the same time, $\mathcal{P}$ are the momenta conjugate to the
coordinates $Q.$ Moreover, the special variables can be chosen such that
$\Omega=\left(  \Omega^{(1)},\Omega^{(2...)}\right)  ,$ where $\Omega^{(1)}$
are primary and$\emph{\ }\Omega^{(2...)}$ are secondary constraints.
Respectively, $\Omega^{(1)}=(\mathcal{P}^{(1)},U^{(1)})$, $\Omega
^{(2...)}=(\mathcal{P}^{(2...)},U^{(2...)})$; $\mathcal{P=}(\mathcal{P}%
^{(1)},\mathcal{P}^{(2...)})$, $U=(U^{(1)},U^{(2...)})$; $\mathcal{P}^{\left(
1\right)  }$ are primary FCC, $\mathcal{P}^{\left(  2...\right)  }$ are
secondary FCC, $U^{\left(  1\right)  }$ are primary SCC, $U^{\left(  2\right)
}$ are secondary SCC. The Hamiltonian action $S_{\mathrm{H}}$ of a general
quadratic gauge theory has the structure
\begin{align}
&  S_{\mathrm{H}}\left[  \mathbf{\vartheta}\right]  =S_{\mathrm{ph}}\left[
\omega\right]  +S_{\mathrm{non-ph}}\left[  \mathbf{\vartheta}\right]
\,,\;\;\mathbf{\vartheta}=\left(  \vartheta,\lambda\right)  \,,\nonumber\\
&  S_{\mathrm{ph}}\left[  \omega\right]  =\int\left[  \omega_{p}\dot{\omega
}_{q}-H_{\mathrm{ph}}\left(  \omega\right)  \right]  dt\,,\nonumber\\
&  S_{\mathrm{non-ph}}\left[  \mathbf{\vartheta}\right]  =\int\left[
\mathcal{P}\dot{Q}+U_{p}\dot{U}_{q}-H_{\mathrm{non-ph}}^{\left(  1\right)
}\left(  \mathbf{\vartheta}\right)  \right]  dt\,, \label{a.1}%
\end{align}
where%
\begin{align}
&  H_{\mathrm{non-ph}}^{\left(  1\right)  }=(Q^{\left(  1\right)
}A+Q^{\left(  2...\right)  }B+\omega C)\mathcal{P}^{\left(  2...\right)
}+\mathcal{P}^{(2...)}D\mathcal{P}^{(2...)}\nonumber\\
&  +\mathcal{P}^{(2...)}EU^{(2...)}+U^{(2...)}GU^{(2...)}+\lambda
_{\mathcal{P}}\mathcal{P}^{(1)}+\lambda_{U}U^{(1)}\,, \label{a.2}%
\end{align}
and $A$, $B$, $C,E$ and $G$ are some matrices (in general case depending on
time). We note that the special variables $\left(  \omega,Q,\Omega\right)  $
may be chosen in more than one way. The equations of motion are%
\[
I\circeq\frac{\delta S_{\mathrm{H}}}{\delta\mathbf{\vartheta}}%
=0\Longrightarrow\left\{
\begin{array}
[c]{c}%
\dot{\vartheta}=\{\vartheta,H^{\left(  1\right)  }\}\\
\Omega=0
\end{array}
\right.  ,
\]
where%
\[
H^{\left(  1\right)  }=H_{\mathrm{ph}}+H_{\mathrm{non-ph}}%
\]
is the total Hamiltonian. In what follows, we call $I$ and $O\left(  I\right)
$ the extremals.

We are going to show that the special phase-space variables can be chosen such
that the non-physical part of the total Hamiltonian (\ref{a.2}) takes a simple
(canonical) form:%
\begin{equation}
H_{\mathrm{non-ph}}^{(1)}=H_{\mathrm{FCC}}^{(1)}+H_{\mathrm{SCC}}^{(1)}\,,
\label{a.3}%
\end{equation}
where%
\begin{align*}
&  H_{\mathrm{FCC}}^{(1)}=\sum_{a=1}^{\aleph_{\chi}}\left(  \sum_{i=1}%
^{a-1}Q^{(i|a)}\mathcal{P}^{(i+1|a)}+\lambda_{\mathcal{P}}^{a}\mathcal{P}%
^{(1|a)}\right)  \,,\\
&  H_{\mathrm{SCC}}^{(1)}=U^{(2...)}FU^{(2...)}+\lambda_{U}U^{(1)}\,.
\end{align*}
Here $\left(  Q,\mathcal{P}\right)  =\left(  Q^{\left(  i|a\right)
},\mathcal{P}^{(i|a)}\right)  \,,\;\lambda_{\mathcal{P}}=\left(
\lambda_{\mathcal{P}}^{a}\right)  \,,\ \;a=1,...,\aleph_{\chi}%
\,,\;i=1,...,a\ ,$ $F$ is a matrix, and $\aleph_{\chi}$ is the number of the
stages of the Dirac procedure that is necessary to determine all the
independent FCC. In what follows, we call such special phase-space variables
the superspecial phase-space variables. In terms of\ the superspecial
phase-space variables, the consistency conditions for the primary FCC
$\mathcal{P}^{\left(  1|a\right)  },$ $a>1$, determine the secondary FCC
$\mathcal{P}^{\left(  2|\aleph_{\chi}\right)  },$ and so on, creating the
following $a$-chain of FCC, $\mathcal{P}^{\left(  1|a\right)  }\rightarrow
\mathcal{P}^{\left(  2|a\right)  }\rightarrow\mathcal{P}^{\left(  3|a\right)
}\cdots\mathcal{P}^{\left(  a|a\right)  },$ see the scheme below,%
\[%
\begin{array}
[c]{ccccccccc}%
\mathcal{P}^{\left(  1|\aleph_{\chi}\right)  } & \rightarrow & \mathcal{P}%
^{\left(  2|\aleph_{\chi}\right)  } & \rightarrow & \cdots & \rightarrow &
\mathcal{P}^{\left(  \aleph_{\chi}-1|\aleph_{\chi}\right)  } & \rightarrow &
\mathcal{P}^{\left(  \aleph_{\chi}|\aleph_{\chi}\right)  }\\
\mathcal{P}^{\left(  1|\aleph_{\chi}-1\right)  } & \rightarrow &
\mathcal{P}^{\left(  2|\aleph_{\chi}-1\right)  } & \rightarrow & \cdots &
\rightarrow & \mathcal{P}^{\left(  \aleph_{\chi}-1|\aleph_{\chi}-1\right)  } &
& \\
\vdots & \vdots & \vdots & \vdots & \vdots &  &  &  & \\
\mathcal{P}^{\left(  1|2\right)  } & \rightarrow & \mathcal{P}^{\left(
2|2\right)  } &  &  &  &  &  & \\
\mathcal{P}^{\left(  1|1\right)  } &  &  &  &  &  &  &  &
\end{array}
\]
The consistency conditions for the constraints $\mathcal{P}^{\left(
a|a\right)  },\;a=1,...,\aleph_{\chi}$ do not yield any new constraints. We
note that in the canonical form the non-physical part of the total Hamiltonian
is independent of the coordinates $Q^{\left(  a|a\right)  }.$

We now prove the above assertion.

First we consider the term $Q^{\left(  1\right)  }A\mathcal{P}^{\left(
2...\right)  }$ in the Hamiltonian (\ref{a.2}). Let the momenta $\mathcal{P}%
^{\left(  1\right)  }$ and the corresponding coordinates $Q^{\left(  1\right)
}$ be labeled by Greek subscripts, while the momenta $\mathcal{P}^{\left(
2...\right)  }$ and the corresponding coordinates $Q^{\left(  2...\right)  }$
be labeled by Latin subscripts,
\[
Q=\left(  Q_{\nu}^{\left(  1\right)  },Q_{b}^{\left(  2...\right)  }\right)
,\;\mathcal{P}=\left(  \mathcal{P}_{\nu}^{\left(  1\right)  },\mathcal{P}%
_{b}^{\left(  2...\right)  }\right)  \,.
\]
We assume that the defect of the rectangular matrix $A^{\nu b}$ is equal to
$\mathbf{a}$ (it is evident that $\left[  Q^{\left(  1\right)  }\right]
-\mathbf{a}\leq\left[  Q^{\left(  2...\right)  }\right]  $). Then, there are
$\mathbf{a}$ nontrivial null vectors $z_{(\tilde{\alpha})},$ $\tilde{\alpha
}=1,...\mathbf{a},$ of the matrix $A$ such that $z_{(\tilde{\alpha})}^{\nu
}A^{\nu b}=0$. We construct a quadratic matrix%
\[
Z_{\alpha}^{\nu}=||z_{(\tilde{\alpha})}^{\nu}z_{(\bar{\alpha})}^{\nu
}||\,,\;\alpha=(\tilde{\alpha},\bar{\alpha})\,,\;\left[  \bar{\alpha}\right]
\leq\left[  b\right]  \,,
\]
where the vectors $z_{(\bar{\alpha})}$ guarantee the nonsingularity of the
complete matrix $Z.$ Such vectors always exist. We then perform the canonical
transformation $\left(  Q^{(1)},\mathcal{P}^{(1)}\right)  \rightarrow\left(
Q^{\prime(1)},\mathcal{P}^{\prime(1)}\right)  ,$ where $Q_{\nu}^{\prime\left(
1\right)  }Z_{\alpha}^{\nu}=Q_{\alpha}^{\left(  1\right)  }.$ Such a canonical
transformation can be performed with a generating function of the form
\begin{equation}
W=Q_{\nu}^{\prime\left(  1\right)  }Z_{\alpha}^{\nu}\mathcal{P}_{\alpha}%
^{(1)}. \label{Fcc.3}%
\end{equation}
We denote the notation
\[
Q_{\alpha}^{\prime(1)}=\left(  Q_{\tilde{\alpha}}^{\prime\left(  1\right)
}=Q_{\tilde{\alpha}}^{\left(  1|1\right)  },Q_{\bar{\alpha}}^{\prime(1)}%
=\bar{Q}_{\bar{\alpha}}^{\left(  1\right)  }\right)  ,\;\mathcal{P}_{\alpha
}^{\prime(1)}=\left(  \mathcal{P}_{\tilde{\alpha}}^{\left(  1|1\right)
},\mathcal{\bar{P}}_{\bar{\alpha}}^{\left(  1\right)  }\right)  \,.
\]
Therefore, the primary FCC now are $\mathcal{P}^{\left(  1|1\right)
},\mathcal{\bar{P}}^{\left(  1\right)  }$ and the corresponding conjugate
coordinates are $Q^{\left(  1|1\right)  },\bar{Q}^{\left(  1\right)  }$. After
the canonical transformation the total Hamiltonian $H^{\left(  1\right)
}=H_{\mathrm{ph}}+H_{\mathrm{non-ph}}$ becomes
\begin{align}
&  H^{\left(  1\right)  }=H_{\mathrm{ph}}+\bar{Q}^{\left(  1\right)
}A^{\prime}\mathcal{P}^{\left(  2...\right)  }+(Q^{\left(  2...\right)
}B+\omega C)\mathcal{P}^{\left(  2...\right)  }+\mathcal{P}^{(2...)}%
D\mathcal{P}^{(2...)}\nonumber\\
&  +\mathcal{P}^{(2...)}EU^{(2...)}+U^{(2...)}FU^{(2...)}+\lambda
_{1}\mathcal{P}^{\left(  1|1\right)  }+\bar{\lambda}\mathcal{\bar{P}}^{\left(
1\right)  }+\lambda_{U}U^{(1)}\,, \label{Fcc.4}%
\end{align}
where%
\[
A^{\prime}=\left(  A^{\prime}\right)  ^{\bar{\nu}b}=Z_{\alpha}^{\bar{\nu}%
}A^{\alpha b}\,,\;\mathrm{rank\,}A^{\prime}=\max=[\bar{Q}^{\left(  1\right)
}];\;b=\left(  \bar{\mu},\bar{b}\right)  \,,\;\det\left(  A^{\prime}\right)
^{\bar{\nu}\bar{\mu}}\neq0\,,
\]
and $\lambda_{1}\mathcal{P}^{\left(  1|1\right)  }+\bar{\lambda}%
\mathcal{\bar{P}}^{\left(  1\right)  }$ denotes the terms proportional to the
primary FCC. At the same, time the functions $\lambda_{1}$ and $\bar{\lambda}$
absorb the time derivative of the generating function (\ref{Fcc.3}). We note
that the coordinates $Q^{(1|1)}$ do not enter the Hamiltonian $H^{\left(
1\right)  }$ (in fact, that was one of the aims of the above canonical
transformation) and therefore, the consistency conditions for the constraints
$\mathcal{P}^{(1|1)}$ do not yield any new constraints,
\[
\left\{  \mathcal{P}^{(1|1)},H^{\left(  1\right)  }\right\}  \equiv0\,.
\]

We consider the consistency conditions for the primary FCC $\mathcal{\bar{P}%
}^{(1)}\,,$%
\[
\left\{  \mathcal{\bar{P}}^{(1)},H^{\left(  1\right)  }\right\}  =-A^{\prime
}\mathcal{P}^{\left(  2...\right)  }=0\,.
\]
Because the rank of the matrix $A^{\prime}$ is maximal, the combinations
$A^{\prime}\mathcal{P}^{\left(  2...\right)  }$ of the secondary FCC are
independent. We can choose them as new momenta $\mathcal{P}^{\prime(2)}$ which
are now second-stage FCC. For this, we perform a canonical transformation
($Q^{(2...)},\mathcal{P}^{(2...)})\rightarrow(Q^{\prime(2...)}$,
$\mathcal{P}^{\prime(2...)}$) with the generating function
\begin{equation}
W=Q^{\prime(2)}A^{\prime}\mathcal{P}^{(2...)}+Q^{\prime(3...)}A^{\prime\prime
}\mathcal{P}^{(2...)}\,. \label{Fcc.5}%
\end{equation}
Here, the rectangular matrix $A^{\prime\prime}$ is chosen such that the
quadratic matrix $\Lambda=||A^{\prime}A^{\prime\prime}||$ is invertible,
$\det\Lambda\neq0$. Therefore, the new variables are
\begin{align*}
&  \mathcal{P}^{\prime(2...)}=\left(  \mathcal{P}^{\prime(2)},\mathcal{P}%
^{\prime(3...)}\right)  ,\;Q^{\prime(2...)}=\left(  Q^{\prime(2)}%
,Q^{\prime(3...)}\right)  \,,\\
&  \mathcal{P}^{\prime(2)}=A^{\prime}\mathcal{P}^{(2...)}\,,\;\mathcal{P}%
^{\prime(3...)}=\left(  A^{\prime\prime}\mathcal{P}^{(2...)}\right)
,\;Q^{\prime(2...)}=Q^{(2...)}\Lambda^{-1}\,.
\end{align*}
In terms of the new variables the Hamiltonian (\ref{Fcc.4}) is%
\begin{align}
&  H^{\left(  1\right)  }=H_{\mathrm{ph}}+\bar{Q}^{\left(  1\right)
}\mathcal{P}^{\prime\left(  2\right)  }+(Q^{\prime\left(  2...\right)
}B^{\prime}+\omega C^{\prime})\mathcal{P}^{\prime\left(  2...\right)
}+\mathcal{P}^{\prime(2...)}D^{\prime}\mathcal{P}^{\prime(2...)}\nonumber\\
&  +\mathcal{P}^{\prime(2...)}E^{\prime}U^{(2...)}+U^{(2...)}FU^{(2...)}%
+\lambda_{1}\mathcal{P}^{\left(  1|1\right)  }+\bar{\lambda}\mathcal{\bar{P}%
}^{\left(  1\right)  }+\lambda_{U}U^{(1)}\,. \label{Fcc.6}%
\end{align}
The matrices $B^{\prime},C^{\prime},D^{\prime},$ and $E^{\prime}$ differ from
$B,C,D,$ and $E$ because of the change of the variables and absorb the time
derivative of the generating function (\ref{Fcc.3}). We note that the latter
derivative does not modify the term $Q^{\prime\left(  1\right)  }%
\mathcal{P}^{\prime\left(  2\right)  }.$

Explicitly separating the terms proportional to $\mathcal{P}^{\prime(2)}$ in
Eqs. (\ref{Fcc.6}) and omitting all the primes, we obtain
\begin{align}
&  H^{\left(  1\right)  }=H_{\mathrm{ph}}+\left(  \bar{Q}^{\left(  1\right)
}+\Sigma_{q}S_{q}+\Sigma_{p}S_{p}\right)  \mathcal{P}^{\left(  2\right)
}+(Q^{\left(  2...\right)  }B+\omega C)\mathcal{P}^{\left(  3...\right)
}\nonumber\\
&  +\mathcal{P}^{(3...)}D\mathcal{P}^{(3...)}+\mathcal{P}^{(3...)}%
EU^{(2...)}+U^{(2...)}FU^{(2...)}+\lambda_{1}\mathcal{P}^{\left(  1|1\right)
}+\bar{\lambda}\mathcal{\bar{P}}^{\left(  1\right)  }+\lambda_{U}U^{(1)}\,,
\label{Fcc.7}%
\end{align}
where $\Sigma=\left(  \Sigma_{q},\Sigma_{p}\right)  $ is the set of all the
phase-space variables except $Q^{\left(  1|1\right)  }$, $\bar{Q}^{(1)}$ and
$\mathcal{P}^{\left(  1|1\right)  }$, $\mathcal{\bar{P}}^{(1)}$, while $S_{q}%
$,$S_{p}$,$B$,$C$,$D,E,$ and $F$ are some matrices.

We now perform a canonical transformation (we do not transform the variables
$Q^{\left(  1|1\right)  },\mathcal{P}^{\left(  1|1\right)  }$) with the
generating function $W$,%
\[
W=\mathcal{\bar{P}}^{\prime(1)}\left(  \bar{Q}^{(1)}+\Sigma_{q}S_{q}%
+\Sigma_{p}^{\prime}S_{p}\right)  +\Sigma_{p}^{\prime}\Sigma_{q}\,,
\]
which yields%
\[
\mathcal{\bar{P}}^{\prime(1)}=\mathcal{\bar{P}}^{(1)}\,,\;\;\bar{Q}%
^{\prime(1)}=\bar{Q}^{(1)}+\Sigma_{q}S_{q}+\Sigma_{p}S_{p}+O(\mathcal{\bar{P}%
}^{(1)}),\;\Sigma^{\prime}=\Sigma+O(\mathcal{\bar{P}}^{(1)})\,.
\]
In terms of the new variables, the Hamiltonian (\ref{Fcc.7}) takes the form
\begin{align}
&  H^{\left(  1\right)  }=H_{\mathrm{ph}}+\bar{Q}^{\left(  1\right)
}\mathcal{P}^{\left(  2\right)  }+(Q^{\left(  2...\right)  }B+\omega
C)\mathcal{P}^{\left(  3...\right)  }+\mathcal{P}^{(3...)}D\mathcal{P}%
^{(3...)}\nonumber\\
&  +\mathcal{P}^{(3...)}EU^{(2...)}+U^{(2...)}FU^{(2...)}+\lambda
_{1}\mathcal{P}^{\left(  1|1\right)  }+\bar{\lambda}\mathcal{\bar{P}}^{\left(
1\right)  }+\lambda_{U}U^{(1)}, \label{Fcc.8}%
\end{align}
where $B$,$C$,$D,E,$ and $F$ are some matrices, the primes are omitted and
redefined functions $\lambda_{\mathcal{P}}$ absorb time derivative of the
generating function.

At this stage of the procedure, we consider the term $Q^{\left(  2\right)
}B\mathcal{P}^{\left(  3...\right)  }$ in the Hamiltonian (\ref{Fcc.8}). Let
the variables $\bar{Q}^{\left(  1\right)  },\mathcal{\bar{P}}^{\left(
1\right)  };$ $Q^{\left(  2\right)  },\mathcal{P}^{\left(  2\right)  }$ be
numbered by Greek subscripts, and the variables $Q^{\left(  3...\right)
},\mathcal{P}^{(3...)}$ be labeled by Latin subscripts (in the general case
the number of indices differs from those from the first stage of the
procedure). We assume that the defect of the rectangular matrix $B^{\nu k}$ is
equal $\mathbf{b}$ (obviously $\left[  Q^{\left(  2\right)  }\right]
-\mathbf{b}\leq\left[  Q^{\left(  3...\right)  }\right]  $). Then, there are
$\mathbf{b}$ nontrivial null vectors $\upsilon_{(\tilde{\alpha})}$,
$\tilde{\alpha}=1,...,\mathbf{b}$ of the matrix $B$ such that $\upsilon
_{(\tilde{\alpha})}^{\nu}B^{\nu k}=0$. We construct a quadratic matrix%
\[
V^{\nu\alpha}=||\upsilon_{(\tilde{\alpha})}^{\nu}\upsilon_{(\bar{\alpha}%
)}^{\nu}||\,,\;\alpha=(\tilde{\alpha},\bar{\alpha})\,,\;[\bar{\alpha}%
]\leq\lbrack k]\,,
\]
where the vectors $\upsilon_{(\bar{\alpha})}$ provide the nonsingularity of
the complete matrix $V.$ Such vectors always exist. Then we perform a
canonical transformation%
\begin{align*}
&  \bar{Q}^{(1)},\mathcal{\bar{P}}^{(1)};\,Q^{(2)},\mathcal{P}^{(2)}%
\rightarrow\bar{Q}^{\prime(1)},\mathcal{\bar{P}}^{\prime(1)};\,Q^{\prime
(2)},\mathcal{P}^{\prime(2)}\,,\\
&  Q^{\prime(2)}=\left(  Q_{\tilde{\alpha}}^{\prime(2)}=Q_{\tilde{\alpha}%
}^{(2|2)},Q_{\bar{\alpha}}^{\prime(2)}=\tilde{Q}_{\bar{\alpha}}^{(2)}\right)
\end{align*}
with the generating function
\[
W=\bar{Q}^{\prime(1)}V\mathcal{\bar{P}}^{(1)}+Q^{\prime(2)}V\mathcal{P}%
^{(2)}\,.
\]
In terms of the new variables, the Hamiltonian (\ref{Fcc.8}) has the form%
\begin{align}
&  H^{\left(  1\right)  }=H_{\mathrm{ph}}+(\bar{Q}^{\left(  1\right)
}+Q^{\left(  2\right)  }\Delta)\mathcal{P}^{(2)}+\tilde{Q}^{\left(  2\right)
}B\mathcal{P}^{\left(  3...\right)  }+(Q^{\left(  3...\right)  }K+\omega
C)\mathcal{P}^{\left(  3...\right)  }\nonumber\\
&  +\mathcal{P}^{(3...)}D\mathcal{P}^{(3...)}+\mathcal{P}^{(3...)}%
EU^{(2...)}+U^{(2...)}FU^{(2...)}+\lambda_{1}\mathcal{P}^{\left(  1|1\right)
}+\bar{\lambda}\mathcal{\bar{P}}^{\left(  1\right)  }+\lambda_{U}U^{(1)}\,,
\label{Fcc.9}%
\end{align}
where $\Delta=\frac{\partial V}{\partial t}V^{-1}$, $B$, $K$, $C$, $D$, $E$,
and $F$ are some matrices (we omit all the primes and redefine $\lambda$). In
particular,%
\[
\mathrm{rank}\,B=\max=\left[  \tilde{Q}^{\left(  2\right)  }\right]
\leq\left[  \mathcal{P}^{\left(  3...\right)  }\right]  \,.
\]

We now perform a canonical transformation $\bar{Q}^{(1)},\mathcal{\bar{P}%
}^{(1)},Q^{(2)},\mathcal{P}^{(2)}\rightarrow\bar{Q}^{\prime(1)},\mathcal{\bar
{P}}^{\prime(1)},Q^{\prime(2)},\mathcal{P}^{\prime(2)}$ with the generating
function
\[
W=\left(  \bar{Q}^{(1)}+Q^{\left(  2\right)  }\Delta\right)  \mathcal{\bar{P}%
}^{\prime(1)}+Q^{(2)}\mathcal{P}^{\prime(2)}\,.
\]
Which yield:%
\[
\mathcal{\bar{P}}^{\prime(1)}=\mathcal{\tilde{P}}^{(1)},\;\bar{Q}^{\prime
(1)}=\bar{Q}^{(1)}+Q^{\left(  2\right)  }\Delta\,,\;\mathcal{P}^{\prime
(2)}=\mathcal{P}^{(2)}-\Delta\mathcal{\bar{P}}^{(1)}\,,\,\;Q^{\prime
(2)}=Q^{(2)}\,.
\]
In terms of the new variables, the Hamiltonian (\ref{Fcc.9}) takes the form%
\begin{align*}
&  H^{\left(  1\right)  }=H_{\mathrm{ph}}+Q^{(1|2)}\mathcal{P}^{(2|2)}%
+\tilde{Q}^{(1)}\mathcal{\tilde{P}}^{(2)}+\tilde{Q}^{(2)}B\mathcal{P}^{\left(
3...\right)  }\\
&  +(Q^{\left(  3...\right)  }K+\omega C)\mathcal{P}^{\left(  3...\right)
}+\mathcal{P}^{(3...)}D\mathcal{P}^{(3...)}+\mathcal{P}^{(3...)}EU^{(2...)}\\
&  +U^{(2...)}FU^{(2...)}+\lambda_{1}\mathcal{P}^{(1|1)}+\lambda
_{2}\mathcal{P}^{(1|2)}+\tilde{\lambda}\mathcal{\tilde{P}}^{\left(  1\right)
}+\lambda_{U}U^{(1)}\,.
\end{align*}
The primes are omitted and $\lambda_{\mathcal{P}}$ are redefined. The time
derivative of the generating function is absorbed by the term $\tilde{\lambda
}\mathcal{\tilde{P}}^{\left(  1\right)  }$.

We note that the variables $Q^{(2|2)}$ do not enter the Hamiltonian and
therefore, the consistency conditions for the constraints $\mathcal{P}%
^{(2|2)}$ do not yield any new constraints. In addition, we remark that at
this stage of the procedure, the primary FCC are $\mathcal{P}^{\left(
1|1\right)  },\mathcal{P}^{(1|2)}$ and $\mathcal{\tilde{P}}^{\left(  1\right)
}.$

We consider the consistency conditions for the second-stage FCC $\mathcal{P}%
^{(2)}\,,$%
\[
\left\{  \mathcal{\tilde{P}}^{(2)},H^{\left(  1\right)  }\right\}
=-B\mathcal{P}^{\left(  3...\right)  }=0\,.
\]
Because the rank of the matrix $B$ is maximum, the combinations $B\mathcal{P}%
^{\left(  3...\right)  }$ are independent. We can choose them as new momenta
$\mathcal{P}^{\prime(3)}$ which are now third-stage FCC. For this, we perform
a canonical transformation ($Q^{(3...)},\mathcal{P}^{(3...)})\rightarrow
(Q^{\prime(3...)}$, $\mathcal{P}^{\prime(3...)}$) with the generating
function
\[
W=Q^{\prime(3)}B\mathcal{P}^{(3...)}+Q^{\prime(4...)}B^{\prime}\mathcal{P}%
^{(3...)}\,,\;Q_{k}^{\left(  3...\right)  }=\left(  Q_{\alpha^{\prime}%
}^{\left(  3\right)  },Q_{k^{\prime}}^{(4...)}\right)  .
\]
The rectangular matrix $B^{\prime}$ is here chosen such that the quadratic
matrix $\Lambda=||BB^{\prime}||$ is invertible, $\det\Lambda\neq0$. We thus,
obtain
\begin{align*}
&  \mathcal{P}^{\prime(3...)}=\Lambda\mathcal{P}=\left(  \mathcal{P}%
_{\alpha^{\prime}}^{\prime(3)},\mathcal{P}_{k^{\prime}}^{\prime(4...)}\right)
\,,\;\mathcal{P}_{\alpha^{\prime}}^{\prime(3)}=B^{\alpha^{\prime}k}%
\mathcal{P}_{k}^{(3...)}\,,\\
&  Q^{\prime(3...)}=Q^{(3...)}\Lambda^{-1}=\left(  Q_{\alpha^{\prime}}%
^{\prime(3)},Q_{k^{\prime}}^{\prime(4...)}\right)  \,.
\end{align*}

In terms of the new variables, the Hamiltonian (\ref{Fcc.9}) has the form%
\begin{align*}
&  H^{\left(  1\right)  }=H_{\mathrm{ph}}+Q^{\left(  1|2\right)  }%
\mathcal{P}^{\left(  2|2\right)  }+\tilde{Q}^{(1)}\mathcal{\tilde{P}}%
^{(2)}+\tilde{Q}^{\left(  2\right)  }\mathcal{P}^{\left(  3\right)
}+(Q^{\left(  3...\right)  }K+\omega C)\mathcal{P}^{\left(  3...\right)
}+\mathcal{P}^{(3...)}D\mathcal{P}^{(3...)}\\
&  +\mathcal{P}^{(3...)}EU^{(2...)}+U^{(2...)}FU^{(2...)}+\lambda
_{1}\mathcal{P}^{\left(  1|1\right)  }+\lambda_{2}\mathcal{P}^{(1|2)}%
+\tilde{\lambda}\mathcal{\tilde{P}}^{\left(  1\right)  }+\lambda_{U}U^{(1)}\,,
\end{align*}
where $K$,$C$,$D,E,$ and $F$ are some matrices and primes are omitted.

We separate terms proportional to $\mathcal{P}^{(3)}$ in this expression and
obtain
\begin{align}
&  H^{\left(  1\right)  }=H_{\mathrm{ph}}+Q^{\left(  1|2\right)  }%
\mathcal{P}^{\left(  2|2\right)  }+\tilde{Q}^{(1)}\mathcal{\tilde{P}}%
^{(2)}+\mathcal{P}^{\left(  3\right)  }\left(  \tilde{Q}^{\left(  2\right)
}+S_{q}\Xi_{q}+S_{p}\Xi_{p}\right) \nonumber\\
&  +(Q^{\left(  3...\right)  }K+\omega C)\mathcal{P}^{\left(  4...\right)
}+\mathcal{P}^{(4...)}D\mathcal{P}^{(4...)}+\mathcal{P}^{(4...)}%
EU^{(2...)}\nonumber\\
&  +U^{(2...)}FU^{(2...)}+\lambda_{1}\mathcal{P}^{\left(  1|1\right)
}+\lambda_{2}\mathcal{P}^{(1|2)}+\tilde{\lambda}\mathcal{\tilde{P}}^{\left(
1\right)  }+\lambda_{U}U^{(1)}\,, \label{Fcc.10}%
\end{align}
where $S_{q},S_{p},K,C,D,E,$and $F$ are some matrices and $\Xi=\left(  \Xi
_{q},\Xi_{p}\right)  $ is the set of all the phase-space variables, except for
$Q^{\left(  1|1\right)  }$, $\mathcal{P}^{\left(  1|1\right)  }$, $Q^{\left(
1|2\right)  }$, $\mathcal{P}^{\left(  1|2\right)  }$, $\tilde{Q}^{(1)}$,
$\mathcal{\tilde{P}}^{(1)}$, $Q^{\left(  2|2\right)  }$, $\mathcal{P}^{\left(
2|2\right)  }$, $\tilde{Q}^{(2)}$, and $\mathcal{\tilde{P}}^{(2)}\,$.

We now perform a canonical transformation (we do not transform the variables
$Q^{\left(  1|1\right)  }$, $\mathcal{P}^{\left(  1|1\right)  }$; $Q^{\left(
1|2\right)  }$, $\mathcal{P}^{\left(  1|2\right)  }$; $\tilde{Q}^{(1)}$,
$\mathcal{\tilde{P}}^{(1)}$, $Q^{\left(  2|2\right)  }$, $\mathcal{P}^{\left(
2|2\right)  }$) with the generating function
\[
W=\mathcal{\tilde{P}}^{\prime(2)}\left(  \tilde{Q}^{(2)}+S_{q}\Xi_{q}+S_{p}%
\Xi_{p}^{\prime}\right)  +\Xi_{p}^{\prime}\Xi_{q}\,.
\]
Which yield
\[
\mathcal{\tilde{P}}^{\prime(2)}=\mathcal{\tilde{P}}^{(2)}\,,\;\tilde
{Q}^{\prime(2)}=\tilde{Q}^{(2)}+S_{q}\Xi_{q}+S_{p}\Xi_{p}\,+O(\mathcal{\tilde
{P}}^{(2)})\,,\;\Xi^{\prime}=\Xi+O(\mathcal{\tilde{P}}^{(2)})\,.
\]
In terms of the new variables, the Hamiltonian (\ref{Fcc.10}) takes the form%
\begin{align}
&  H^{\left(  1\right)  }=H_{\mathrm{ph}}+Q^{\left(  1|2\right)  }%
\mathcal{P}^{\left(  2|2\right)  }+\mathcal{\tilde{P}}^{(2)}\left(  \tilde
{Q}^{(1)}+R_{q}\Sigma_{q}+R_{p}\Sigma_{p}\right)  +\tilde{Q}^{\left(
2\right)  }\mathcal{P}^{\left(  3\right)  }\nonumber\\
&  +(Q^{\left(  3...\right)  }K+\omega C)\mathcal{P}^{\left(  4...\right)
}+\mathcal{P}^{(4...)}D\mathcal{P}^{(4...)}+\mathcal{P}^{(4...)}%
EU^{(2...)}\nonumber\\
&  +U^{(2...)}FU^{(2...)}+\lambda_{1}\mathcal{P}^{\left(  1|1\right)
}+\lambda_{2}\mathcal{P}^{(1|2)}+\tilde{\lambda}\mathcal{\tilde{P}}^{\left(
1\right)  }+\lambda_{U}U^{(1)}\,, \label{Fcc.11}%
\end{align}
where $\Sigma=(\tilde{Q}^{(2)},\mathcal{\tilde{P}}^{(2)},\Xi)=\left(
\Sigma_{q},\Sigma_{p}\right)  $ and $R,K,C,D,E,$ and $F$ are some matrices,
all the primes are omitted.

We perform a canonical transformation with the generating function%
\[
W=\mathcal{\tilde{P}}^{\prime(1)}\left(  \tilde{Q}^{(1)}+R_{q}\Sigma_{q}%
+R_{p}\Sigma_{p}^{\prime}\right)  +\Sigma_{p}^{\prime}\Sigma_{q}\,,
\]
and obtain
\[
\mathcal{\tilde{P}}^{\prime(1)}=\mathcal{\tilde{P}}^{(1)}\,,\;\tilde
{Q}^{\prime(1)}=\tilde{Q}^{(1)}+R_{q}\Sigma_{q}+R_{p}\Sigma_{p}%
\,+O(\mathcal{\tilde{P}}^{(1)})\,,\;\Sigma^{\prime}=\Sigma+O(\mathcal{\tilde
{P}}^{(1)})\,.
\]
In terms of the new variables, the Hamiltonian (\ref{Fcc.11}) takes the form
\begin{align}
&  H^{\left(  1\right)  }=H_{\mathrm{ph}}+Q^{\left(  1|2\right)  }%
\mathcal{P}^{\left(  2|2\right)  }+\tilde{Q}^{(1)}\mathcal{\tilde{P}}%
^{(2)}+\tilde{Q}^{\left(  2\right)  }\mathcal{P}^{\left(  3\right)
}\nonumber\\
&  +(Q^{\left(  3...\right)  }K+\omega C)\mathcal{P}^{\left(  4...\right)
}+\mathcal{P}^{(4...)}D\mathcal{P}^{(4...)}+\mathcal{P}^{(4...)}%
EU^{(2...)}\nonumber\\
&  +U^{(2...)}FU^{(2...)}+\lambda_{1}\mathcal{P}^{\left(  1|1\right)
}+\lambda_{2}\mathcal{P}^{(1|2)}+\tilde{\lambda}\mathcal{\tilde{P}}^{\left(
1\right)  }+\lambda_{U}U^{(1)}\,, \label{Fcc.12}%
\end{align}
where $K,C,D,E,$and $F$ are some matrices, all the primes are omitted and
$\lambda_{\mathcal{P}}$ are redefined.

Further transformations of the Hamiltonian (\ref{Fcc.12}) can be done using
the same kind of canonical transformations as the ones used before. In the end
of the procedure, we obtain at the form (\ref{a.3}) for the non-physical part
of the total Hamiltonian.

We emphasize some important facts related to the canonical transformation that
was performed to reduce the total Hamiltonian to the form (\ref{a.3}).

First, we note that the final variables $\omega$, $Q$, $\Omega,$ where
$\Omega=(\mathcal{P}$, $U)$ (superspecial phase-space variables) still remain
special phase-space canonical variables $\vartheta$ and possess all the
corresponding properties of such variables. Let the final superspecial
phase-space canonical variables are labeled by primes while the initial
special phase-space variables are without primes. We can see that
\begin{align*}
\mathcal{P}^{\prime}  &  =T\mathcal{P\,},\;\mathcal{P}^{(1)\prime}=T^{\left(
1\right)  }\mathcal{P}^{(1)}\,,\\
U^{\prime}  &  =U+O(\mathcal{P})\,,\;U^{(1)\prime}=U^{\left(  1\right)  }\,,
\end{align*}
such that $\mathcal{P}^{\prime}$ are FCC, $\mathcal{P}^{(1)\prime}$ are
primary FCC, $U^{\prime}$ are SCC, and $U^{(1)\prime}$ are primary SCC. The
physical variables do not change on the constraint surface, $\omega
\rightarrow\omega^{\prime}=\omega+O(\mathcal{P})$. We emphasize that the
superspecial variables $\mathcal{P}^{(i|a)}$ coincide with the FCC
$\chi^{\left(  i|a\right)  }$ in the orthogonal constraint basis introduced in
\cite{196}. In the general nonquadratic theory, the relation is
\begin{equation}
\chi^{\left(  i|a\right)  }=\mathcal{P}^{(i|a)}+O\left(  \vartheta
\Omega\right)  \,. \label{Fcc.13}%
\end{equation}

We can also see that in the superspecial phase-space variables, the
non-physical part of the Hamiltonian action can be written as:%
\begin{equation}
S_{\mathrm{non-ph}}=\int\left[  \mathcal{P}\hat{\Lambda}\mathcal{Q}+\sum
_{i=1}^{\aleph_{\chi}}\mathcal{P}^{(i|i)}\dot{Q}^{(i|i)}+\mathcal{U}\hat
{B}\mathcal{U}\right]  dt\,, \label{Fcc.15}%
\end{equation}
where $\hat{\Lambda}$ and $\hat{B}$ are first-order differential matrix
operators and%
\[
\mathcal{Q}=(\lambda_{\mathcal{P}}^{a},\,Q^{(i|a)}%
\,,\;\,i=1,...,a-1\,,\;a=1,...,\aleph_{\chi})\,,\;\mathcal{U}=(\lambda
_{U},U)\,.
\]
It is important that $\left[  \mathcal{Q}\right]  =\left[  \mathcal{P}\right]
\,$because $\left[  \lambda_{\mathcal{P}}\right]  =\left[  \mathcal{P}%
^{(1)}\right]  .$

We can see that there are local operators $\hat{\Lambda}^{-1}$ and $\hat
{B}^{-1}$ such that $\hat{\Lambda}\hat{\Lambda}^{-1}=\hat{\Lambda}^{-1}%
\hat{\Lambda}=1$, $\hat{B}\hat{B}^{-1}=\hat{B}^{-1}\hat{B}=1$. This assertion
can be derived from the fact that by the construction of the special
phase-space variables, the Hamiltonian equations of motion have the unique
solution $\mathcal{P}=0$ and $\mathcal{U}=0$. Therefore, the equations%
\begin{equation}
\frac{\delta S_{\mathrm{H}}}{\delta\mathcal{Q}}=0\Longrightarrow\hat{\Lambda
}^{T}\mathcal{P}=0\,,\;\frac{\delta S_{\mathrm{H}}}{\delta\mathcal{U}%
}=0\Longrightarrow\hat{B}\mathcal{U}=0 \label{Fcc.16}%
\end{equation}
must have only the solution $\mathcal{P}=0$ and $\mathcal{U}=0$. We represent
$\hat{\Lambda}$ as%
\[
\hat{\Lambda}=\Lambda\left(  \frac{d}{dt}\right)  =a\frac{d}{dt}+b\,,
\]
where $a$ and $b$ are some constant matrices, and consider the solutions of
the form $\mathcal{P}(t)=e^{-Et}\,\mathcal{P}(0)\,,$ where $E$ is a complex
number. We obtain that $\Lambda^{T}(E)\mathcal{P}(0)=0.$ The existence of the
unique solution $\mathcal{P}(0)=0$ implies%
\begin{equation}
\forall E:\;\det\Lambda(E)\neq0\,. \label{Fcc.17}%
\end{equation}
On the other hand, $\det\Lambda(E)$ is a polynomial of $E$. Because of
(\ref{Fcc.17}), such a polynomial has no roots. That means that $\det
\Lambda(E)=\mathrm{const}=c.$ In turn, this implies that%
\[
\Lambda^{-1}(E)=\frac{1}{c}\Delta(E)\,,
\]
where $\Delta(E)$ are the corresponding minors of the matrix $\Lambda(E)$. The
latter minors are finite order polynomials in $E$. Therefore, the operator%
\[
\hat{\Lambda}^{-1}=\frac{1}{c}\Delta\left(  \frac{d}{dt}\right)
\]
is a local operator. We can similarly prove the existence of the local
operator $\hat{B}^{-1}$ (to this, it is convenient to reduce the Hamiltonian
$H_{\mathrm{SCC}}^{(1)}$ to a canonical form, see below).

\section{Functional arbitrariness in solutions of equations of motion}

In theories with FCC, the equations of motion do not determine a unique
trajectory for given initial data. In what follows, we study this problem for
the quadratic gauge theories using the superspecial phase-space variables. The
equations of motion that follow from the action (\ref{a.1}) and (\ref{a.2}),
with taking (\ref{a.3}) into account, are%
\begin{equation}
\dot{\omega}=\{\omega,H_{\mathrm{ph}}\}\,,\;\Omega=0\,, \label{c.1}%
\end{equation}
and%
\begin{equation}
\dot{Q}^{(i|a)}=\{Q^{(i|a)},H_{\mathrm{non-ph}}\}\Longrightarrow\left\{
\begin{array}
[c]{c}%
\dot{Q}^{\left(  1|a\right)  }=\lambda_{\mathcal{P}}^{a}\,,\\
\dot{Q}^{\left(  2|a\right)  }=Q^{(1|a)}\,,\\
....\\
\dot{Q}^{\left(  a|a\right)  }=Q^{(a-1|a)}\,.
\end{array}
\right.  \label{c.2}%
\end{equation}

We see that the equations (\ref{c.1}) for the physical variables $\omega$ and
for $\Omega$ have a unique solution whenever initial data for these variables
are given. There exists a functional arbitrariness in solutions of the
equations of motion (\ref{c.2}) for the variables $Q$, because these equations
contain arbitrary functions of time $\lambda_{\mathcal{P}}\left(  t\right)  .$
We note that the number of variables $Q$ is equal to the number of all FCC
and, in general, this number is larger than the number of the arbitrary
functions $\lambda_{\mathcal{P}}\left(  t\right)  $. However, as it will be
seen below, because of the specific structure of the equations, the
''influence'' of these arbitrary functions on solutions for $Q$ is very
strong. This fact is extremely important for the physical interpretation of
the variables $Q$ and for the general physical interpretation of theories with
FCC. The following proposition describes to what extent the variables $Q$ are
affected by the arbitrary functions $\lambda_{\mathcal{P}}\left(  t\right)  $.

The equations of motion (\ref{c.2}) for the variables $Q$ are completely
controllable\footnote{For exact definition of the controllability see e.g. the
book \cite{LeeMa67}.} by the functions $\lambda_{\mathcal{P}}\left(  t\right)
.$ In the case under consideration, this means that under a proper choice of
the functions $\lambda_{\mathcal{P}}^{a}\left(  t\right)  ,$ the equations
(\ref{c.2}) have a solution with the properties%
\begin{align}
&  \left.  Q^{(i|a)}\right|  _{t=0}=0\,,\;\;\left.  Q^{(i|a)}\right|
_{t=\tau}=\Delta^{(i|a)},\;i=1,...,a\,,\nonumber\\
&  \left.  \frac{d^{s}\lambda_{\mathcal{P}}^{a}}{d^{s}t}\right|
_{t=0}=0\,,\;\;\left.  \frac{d^{s}\lambda_{\mathcal{P}}^{a}}{d^{s}t}\right|
_{t=\delta}=\delta_{(s)}^{a}\,,\;s=0,1,...,K\,\,, \label{c.3}%
\end{align}
where $\tau,\;\Delta^{(i|a)}$, $\Delta^{(i|a)}$, $\delta_{(s)}^{a},\;$and the
integer $K$ are arbitrary.

Because of a simple structure of the equations of motion in superspecial
phase-space variables, the proof of the above assertion can be done in a
constructive manner. Namely, we explicitly present such a solution. It has the
form%
\[
Q^{\left(  i|a\right)  }=\frac{d^{a-i}X^{a}}{dt^{a-i}}\,,\;i=1,...a\,,
\]
if we choose%
\[
\lambda_{\mathcal{P}}^{a}=\frac{d^{a}X^{a}}{dt^{a}}\,,
\]
where $X^{a}\left(  t\right)  $ are arbitrary smooth functions obeying the
following boundary conditions%
\begin{align*}
&  \left.  \frac{d^{s}X^{a}}{dt^{s}}\right|  _{t=0}=0\,,\;s=0,...,K+a\,,\\
&  \left.  \frac{d^{s}X^{a}}{dt^{s}}\right|  _{t=\tau}=\left\{
\begin{array}
[c]{c}%
\left.  Q^{(a-s|a)}\right|  _{t=\tau}=\Delta^{(a-s|a)}\,,\;s=0,...,a-1\,,\\
\left.  \frac{d^{s-a}\lambda_{\mathcal{P}}^{a}}{dt^{s-a}}\right|  _{t=\tau
}=\delta_{(s-a)}^{a}\,,\;s=a,...,K+a\,.
\end{array}
\right.
\end{align*}
For example, the functions $X^{a}\left(  t\right)  $ can be chosen as:%
\[
X^{a}\left(  t\right)  =f(t)\left[  \sum_{s=0}^{a-1}\frac{1}{s!}%
\Delta^{(a-s|a)}(t-\delta)^{s}+\sum_{s=a}^{K+a}\frac{1}{s!}\delta
_{(s-a+1)}^{a}(t-\delta)^{s}\right]
\]
where $f(t)$ is an arbitrary smooth function that is respectively equal to
zero and to one in the neighborhoods of the points $t=0$ and $t=\tau$. An
example of such a function is \footnote{Here and in what follows, we use the
notation%
\[
f^{\left[  s\right]  }=\frac{d^{s}f}{dt^{s}}\,.
\]
}%
\begin{align}
&  f(t)=\left\{
\begin{array}
[c]{c}%
0\,,\;\;\;\;\;t\leq\varepsilon\,,\\
\frac{1}{1+e^{u}}\,,\;u=\frac{1}{t-\varepsilon}+\frac{1}{t-\left(
\tau-\varepsilon\right)  }\,,\;\varepsilon\leq t\leq\tau-\varepsilon\,,\\
1\,,\;\;\;\;\;t\geq\tau-\varepsilon\,,
\end{array}
\right. \nonumber\\
&  \lim_{t\rightarrow\varepsilon+0}\;f^{\left[  s\right]  }(t)=0\,,\;\;\lim
_{t\rightarrow\tau-\varepsilon-0}\;f^{\left[  s\right]  }(t)=\delta
_{0,s}\,,\;\;s\geq0\,. \label{c.4}%
\end{align}

The proved proposition is crucial for the understanding of the structure of
theories with FCC (gauge theories) and for their physical interpretation. The
most remarkable fact is the following: The functional arbitrariness in
equations of motion of theories with FCC (gauge theories) is due to the
undetermined Lagrange multipliers to the primary FCC. However, this
arbitrariness affects essentially more variables. In the special variables,
all the variables $Q$ are controllable by the undetermined Lagrange
multipliers. The number of $Q$-variables is equal to the number of all the FCC
and is greater than the number of the Lagrange multipliers.

\section{Symmetries}

We recall that a transformation $q\left(  t\right)  \rightarrow q^{\prime
}\left(  t\right)  $ is called a symmetry of an action $S$ if%
\begin{equation}
L\left(  q,\dot{q}\right)  \rightarrow L^{\prime}\left(  q,\dot{q}\right)
=L\left(  q,\dot{q}\right)  +\frac{dF}{dt}\,,\label{I.5}%
\end{equation}
where $F$ is a local function. In what follows, only infinitesimal symmetry
transformations $q\rightarrow q+\delta q$ with local finctions $\delta q$ are
considered. These symmetry transformations can be global, gauge, and trivial
ones. Gauge transformations are parametrized by some arbitrary functions of
time, gauge parameters (in the case of a field theory the gauge parameters
depend on all space-time variables). Any infinitesimal symmetry transformation
implies a conservation low \textbf{(}N\"{o}ether theorem):
\begin{align}
&  \frac{dG}{dt}=-\delta q^{a}\frac{\delta S}{\delta q^{a}}\Longrightarrow
G=\mathrm{const.\;on\;extremals,}\label{I.11}\\
&  G=P-F\,,\;P=\frac{\partial L}{\partial\dot{q}^{a}}\delta q^{a},\;\delta
L=\frac{dF}{dt}\,.\nonumber
\end{align}
The local function $G$ is referred to as the conserved charge related to the
symmetry $\delta q$ of the action $S$. The quantities $\delta q,\;S,$ and $G$
are related by the equation (\ref{I.11}). In what follows, we call this
equation the symmetry equation. The symmetry equation for the Hamiltonian
action $S_{\mathrm{H}}\left[  \mathbf{\vartheta}\right]  $ has the form%
\begin{equation}
\delta\mathbf{\vartheta}\frac{\delta S_{\mathrm{H}}}{\delta\mathbf{\vartheta}%
}+\frac{dG}{dt}=0\,.\label{I.12}%
\end{equation}

\subsection{Trivial symmetries}

For any action, there are trivial symmetry transformations,
\begin{equation}
\delta_{\mathrm{tr}}q^{a}=\hat{U}^{ab}\frac{\delta S}{\delta q^{b}}\,,
\label{I.9}%
\end{equation}
where $\hat{U}$ is an antisymmetric local operator, that is $\left(
\hat{U}^{T}\right)  ^{ab}=-\hat{U}^{ab}$ . The trivial symmetry
transformations do not affect genuine trajectories.

Using the simple action structure in superspecial phase-space variables, we
can prove the following assertion: for theories with FCC, symmetries of the
Hamiltonian action that vanish on the extremals are trivial symmetries.

To prove this assertion we consider the Hamiltonian action $S_{\mathrm{H}%
}\left[  \mathbf{\vartheta}\right]  ,\;\mathbf{\vartheta}=\left(
\vartheta,\lambda\right)  $ of a theory with FCC in the superspecial
phase-space variables $\vartheta$. We can see that the equations of motion
\[
\frac{\delta S_{\mathrm{H}}}{\delta\mathcal{U}}=0\,,\;\frac{\delta
S_{\mathrm{H}}}{\delta\mathcal{Q}}=0\,,\;\frac{\delta S_{\mathrm{H}}}%
{\delta\mathcal{P}}=0
\]
have solution of the form $\mathcal{U}=\mathcal{P}=0$, $\mathcal{Q}%
=\psi\left(  Q^{\left(  i|i\right)  }\right)  ,$ where $\psi$ are local
functions of the indicated arguments. Therefore, the variables $\mathcal{U}%
,\mathcal{P}$, and$\;\mathcal{Q}$ are auxiliary ones\footnote{Suppose an
action $S[q,y]$ contains two groups of coordinates $q$ and $y$ such that the
coordinates $y$ can be expressed as local functions $y=\bar{y}\left(
q^{\left[  l\right]  },l<\infty\right)  \,$of $q$ and their time derivatives
by the help of the equations $\delta S/\delta y=0$. We call $y$ the auxiliary
coordinates. The action $S[q,y]$ and the reduced action $S\left[  q\right]
=S[q,\bar{y}]$ lead to the same equations for the coordinates $q,$ see
\cite{Superflu,BorTy98}. The actions $S[q,y]$ and $S\left[  q\right]  $ are
called dynamically equivalent actions. One can prove that there exists a
one-to-one correspondence (isomorphism) between the symmetry classes of the
extended action. Symmetries are equivalent if they differ by a trivial
transformation.}. Excluding these variables from the action $S_{\mathrm{H}},$
we obtain a dynamically equivalent action $\bar{S}_{\mathrm{H}}\left[
\omega,Q^{\left(  i|i\right)  }\right]  $. Taking into account that
$\mathcal{U}=\mathcal{P}=0\Longrightarrow\Omega=0,$ and the relation $\left.
S_{\mathrm{H}}\right|  _{\Omega=0}=S_{\mathrm{ph}}\left[  \omega\right]  $, we
find%
\[
\bar{S}_{\mathrm{H}}\left[  \omega,Q^{\left(  i|i\right)  }\right]  =\left.
S_{\mathrm{H}}\right|  _{\mathcal{U}=\mathcal{P}=0,\mathcal{Q}=\psi
}=S_{\mathrm{ph}}\left[  \omega\right]  \,.
\]

Let a transformation $\delta\mathbf{\vartheta}$ which vanishes on the
extremals be a symmetry of the action $S_{\mathrm{H}}$ . Consider the reduced
transformation $\bar{\delta}\omega$, $\bar{\delta}Q^{\left(  i|i\right)  },$%
\[
(\bar{\delta}\omega,\bar{\delta}Q^{\left(  i|i\right)  })=\left.
\delta\mathbf{\vartheta}\right|  _{\mathcal{U}=\mathcal{P}=0,\mathcal{Q}=\psi
}\,.
\]
It is evident that the reduced transformation vanishes on the extremals of the
reduced action $\bar{S}_{\mathrm{H}}$ and is a symmetry transformation of the
action $\bar{S}_{\mathrm{H}}$. This implies that%
\[
\bar{\delta}\omega=\hat{m}\frac{\delta S_{\mathrm{ph}}}{\delta\omega}%
\,,\;\bar{\delta}Q^{\left(  i|i\right)  }=\left(  \hat{n}\frac{\delta
S_{\mathrm{ph}}}{\delta\omega}\right)  ^{\left(  i|i\right)  }\,,
\]
where $\hat{m}$ and $\hat{n}$ are some local operators. The transformation
$\bar{\delta}\omega$ is obviously the symmetry transformation of the
nonsingular action $S_{\mathrm{ph}}$, and in addition, this symmetry
transformation vanishes on the extremals. We can prove that such a symmetry is
always a trivial one and, therefore, $\hat{m}$ is antisymmetric. Therefore,
the complete transformation $\bar{\delta}\omega$, $\bar{\delta}Q^{\left(
i|i\right)  }$ can be represented as
\[
\left(
\begin{array}
[c]{c}%
\bar{\delta}\omega\\
\bar{\delta}Q^{\left(  |\right)  }%
\end{array}
\right)  =\hat{M}\left(
\begin{array}
[c]{c}%
\frac{\delta\bar{S}_{\mathrm{H}}}{\delta\omega}\\
\frac{\delta\bar{S}_{\mathrm{H}}}{\delta Q^{\left(  |\right)  }}%
\end{array}
\right)  \,,\;\hat{M}=\left(
\begin{array}
[c]{cc}%
\hat{m} & -\hat{n}^{T}\\
\hat{n} & 0
\end{array}
\right)  \,.
\]
The matrix $\hat{M}$ is evidently antisymmetric.

Finally, the transformation $\bar{\delta}\omega$, $\bar{\delta}Q^{\left(
i|i\right)  }$ is a trivial symmetry of the action $\bar{S}_{\mathrm{H}}.$
This implies that the extended transformation $\delta\mathbf{\vartheta}$ is a
trivial symmetry of the action $S_{\mathrm{H}}$.

\subsection{Gauge symmetries}

We are now going to prove the following assertion:

In theories with FCC, there exist nontrivial symmetries $\delta
\mathbf{\vartheta}$ of the Hamiltonian action $S_{\mathrm{H}}$ that are gauge
transformations. These symmetries are parametrized by the gauge parameters
$\nu$. These parameters are arbitrary functions of time $t$. Moreover, they
can be arbitrary local functions of $\mathbf{\vartheta}=(\vartheta,\lambda)$.

The corresponding conserved charge (the gauge charge) is a local
function\footnote{A local function depends on some variables $x^{a\left[
l\right]  },\;a=1,...,n,\;l=0,1,...,N$ up to some finite order $N$.\ We use
the following notation for the local functions:
\[
F\left(  x^{a},x^{a\left[  1\right]  },x^{a\left[  2\right]  },...\right)
=F\left(  x^{\left[  {}\right]  }\right)  \,.
\]
} $G=G\left(  \mathcal{P},\nu^{\lbrack]}\right)  $, which vanishes on the
extremals. The gauge charge has the following decomposition with respect to
the FCC:%
\begin{equation}
G=\sum_{i=1}^{\aleph_{\chi}}\nu_{\left(  a\right)  }\mathcal{P}^{(a|a)}%
+\sum_{i=1}^{\aleph_{\chi}-1}\sum_{a=i+1}^{\aleph_{\chi}}C_{i|a}%
\mathcal{P}^{(i|a)}\,. \label{sdc.4}%
\end{equation}
Here $C_{i|a}\left(  \nu^{\lbrack]}\right)  $ are some local functions, which
can be determined from the symmetry equation in an algebraic way, and
$\nu=\left(  \nu_{\left(  a\right)  }\right)  ,\;\nu_{\left(  a\right)
}=\left(  \nu_{\left(  a\right)  }^{\mu_{a}}\right)  ,\;a=1,...,\aleph_{\chi
}.$ Here $\nu_{\left(  a\right)  }$ are gauge parameters related to the FCC in
a chain whose number is $a.$ The number of the gauge parameters $\nu_{\left(
a\right)  }$ is equal to the number of the primary FCC in the chain $a.$ The
index $\mu_{a}$ labels constraints (and gauge parameters) inside the chain.

The total number of the gauge parameters is equal to the number of the primary
FCC. The total number of the independent gauge parameters together with their
time derivatives that enter essentially in the gauge charge is equal to the
number of all the FCC.

The gauge charge is the generating function for the variations $\delta
\vartheta$ of the phase-space variables,%
\begin{equation}
\delta\vartheta=\left\{  \vartheta,G\right\}  \,. \label{sdc.7}%
\end{equation}
The variations $\delta\lambda_{\mathcal{U}}$ are vanishing, and $\delta
\lambda_{\mathcal{P}}^{a}=\nu_{(a)}^{[a]}$.

To prove the above assertion, we consider the symmetry equation (\ref{I.11})
for the case under consideration. Taking the action structure (\ref{a.1}%
,\ref{a.3}), the anticipated form of the gauge charge (\ref{sdc.4}), and of
the variations $\delta\vartheta,$ into account, we can rewrite this equation
as follows:%
\begin{equation}
\hat{H}G-\sum_{a=1}^{\aleph_{\chi}}\lambda_{\mathcal{P}}^{a}\left\{
\mathcal{P}^{(1\left|  a\right|  )},G\right\}  =U^{\left(  1\right)  }%
\delta\lambda_{U}+\sum_{a=1}^{\aleph_{\chi}}\delta\lambda_{\mathcal{P}}%
^{a}\mathcal{P}^{(1|a)}\,, \label{sdc.9a}%
\end{equation}
where%
\begin{align}
&  \hat{H}G=\left\{  G,H\right\}  +\left(  \frac{\partial}{\partial t}%
+\nu^{\left[  m+1\right]  }\frac{\partial}{\partial\nu^{\left[  m\right]  }%
}\right)  G\,,\nonumber\\
&  H=H_{\mathrm{ph}}(\omega)+\sum_{a=1}^{\aleph_{\chi}}\sum_{i=1}%
^{a-1}Q^{(i|a)}\mathcal{P}^{(i+1|a)}+U^{(2...)}FU^{(2...)}\,. \label{sdc.10a}%
\end{align}
The following commutation relations%
\begin{align*}
&  \left\{  \mathcal{P}^{(i|a)},H\right\}  =-\mathcal{P}^{(i+1|a)}%
\,,\;i=1,...,\aleph_{\chi}-1\,,\;a=i+1,...,\aleph_{\chi}\,,\\
&  \left\{  \mathcal{P}^{\left(  i|i\right)  },H\right\}
=0\,,\;i=1,...,\aleph_{\chi}\,;\;\left\{  \mathcal{P},\Omega\right\}  =0\,
\end{align*}
hold. The equation (\ref{sdc.9a}) implies the following equations for the
functions $C_{i|a}$ and for the variations $\delta\lambda$
\begin{align}
&  \sum_{i=1}^{\aleph_{\chi}-1}\sum_{a=i+1}^{\aleph_{\chi}}\left[
-\mathcal{P}^{(i+1|a)}+\mathcal{P}^{(i|a)}\left(  \frac{\partial}{\partial
t}+\nu^{\left[  m+1\right]  }\frac{\partial}{\partial\nu^{\left[  m\right]  }%
}\right)  \right]  C_{i|a}\nonumber\\
&  +\sum_{a=1}^{\aleph_{\chi}}\mathcal{P}^{(a|a)}\dot{\nu}_{(a)}=U^{\left(
1\right)  }\delta\lambda_{U}+\sum_{a=1}^{\aleph_{\chi}}\mathcal{P}%
^{(1|a)}\delta\lambda_{\mathcal{P}}^{a}\,. \label{sdc.151}%
\end{align}
Considering Eq. (\ref{sdc.151}) on the constraint surface $\mathcal{P}%
^{(i|a)}=0$, $i=1,...,\aleph_{\chi}-1$, $a=i,...,\aleph_{\chi}\,,\;U^{\left(
1\right)  }=0$, we can choose $C_{\aleph_{\chi}-1|\aleph_{\chi}}=\dot{\nu
}_{(\aleph_{\chi})}\,.$ Substituting this $C_{\aleph_{\chi}-1|\aleph_{\chi}}$
into Eq. (\ref{sdc.151}), we obtain that%
\begin{align}
&  \sum_{i=1}^{\aleph_{\chi}-2}\sum_{a=i+1}^{\aleph_{\chi}}\left[
-\mathcal{P}^{(i+1|a)}+\mathcal{P}^{(i|a)}\left(  \frac{\partial}{\partial
t}+\nu^{\left[  m+1\right]  }\frac{\partial}{\partial\nu^{\left[  m\right]  }%
}\right)  \right]  C_{i|a}\nonumber\\
&  +\mathcal{P}^{(\aleph_{\chi}-1|\aleph_{\chi})}\nu_{\aleph_{\chi}}%
^{[2]}+\sum_{a=1}^{\aleph_{\chi}-1}\mathcal{P}^{(a|a)}\dot{\nu}_{(a)}%
=U^{\left(  1\right)  }\delta\lambda_{U}+\sum_{a=1}^{\aleph_{\chi}}%
\mathcal{P}^{(1|a)}\delta\lambda_{\mathcal{P}}^{a}\,. \label{sdc.161}%
\end{align}
Considering the equation (\ref{sdc.161}) on the constraint surface
$\mathcal{P}^{(i|a)}=0$, $i=1,...,\aleph_{\chi}-2$, $a=i,...,\aleph_{\chi
}\,,\;U^{\left(  1\right)  }=0$, we choose $C_{\aleph_{\chi}-2|\aleph_{\chi
}-1}=\dot{\nu}_{(\aleph_{\chi}-1)}\,,\;C_{\aleph_{\chi}-2|\aleph_{\chi}}%
=\ddot{\nu}_{(\aleph_{\chi})}\,.$ We see that we can similarly determine all
the $C_{i|a}$ such that
\begin{equation}
C_{i|a}=\nu_{(a)}^{[a-i]},\;i=1,...,\aleph_{\chi}-1,\;a=i+1,...,\aleph_{\chi
}\,. \label{sdc.171}%
\end{equation}
Therefore, in the case under consideration, the gauge charge has the following
form%
\begin{equation}
G=\sum_{i=1}^{\aleph_{\chi}}\sum_{a=i}^{\aleph_{\chi}}\nu_{(a)}^{[a-i]}%
\mathcal{P}^{(i|a)}\,. \label{sdc.181}%
\end{equation}

The form of the variations $\delta\vartheta$ follows from (\ref{sdc.7}),%
\begin{equation}
\delta Q^{(i|a)}=\nu_{(a)}^{[a-i]},\;\delta\omega=\delta\Omega=0\,.
\label{sdc.201}%
\end{equation}

After all the $C_{i|a}$ are known, the variations $\delta\lambda$ can be
determined from Eq. (\ref{sdc.151}),
\begin{equation}
\delta\lambda_{U}=0\,,\;\delta\lambda_{\mathcal{P}}^{a}=\nu_{(a)}^{[a]}\,,
\label{sdc.211}%
\end{equation}

which proves the assertion.

\section{Structure of arbitrary symmetry}

Analyzing the symmetry equation, we are going to prove that:

Any symmetry $\delta\mathbf{\vartheta}$ and $G$ of the action $S_{\mathrm{H}}$
can be presented as the sum of three type of symmetries%
\begin{equation}
\left(
\begin{array}
[c]{c}%
\delta\mathbf{\vartheta}\\
G
\end{array}
\right)  =\left(
\begin{array}
[c]{c}%
\delta_{c}\mathbf{\vartheta}\\
G_{c}%
\end{array}
\right)  +\left(
\begin{array}
[c]{c}%
\delta_{g}\mathbf{\vartheta}\\
G_{g}%
\end{array}
\right)  +\left(
\begin{array}
[c]{c}%
\delta_{\mathrm{tr}}\mathbf{\vartheta}\\
G_{\mathrm{tr}}%
\end{array}
\right)  \,, \label{sdc.24}%
\end{equation}
such that:

The set $\delta_{c}\mathbf{\vartheta}$ and $G_{c}$ is a global symmetry,
canonical for the phase-space variables $\vartheta$. All the variations
$\delta_{c}\mathbf{\vartheta}$ and the corresponding conserved charge $G_{c}$
are either identically zero or do not vanish on the extremals.

The set $\delta_{g}\mathbf{\vartheta}$ and $G_{g}$ is a particular gauge
transformation given by Eqs. (\ref{sdc.181}), (\ref{sdc.201}), and
(\ref{sdc.211}) with specific fixed gauge parameters (i.e., specific fixed
forms of the functions $\nu=\bar{\nu}\left(  t,\eta^{\lbrack]},\lambda
^{\lbrack]}\right)  $) that are either identically zero or do not vanish on
the extremals. In the latter case, the corresponding conserved charge $G_{g}$
vanishes on the extremals, whereas the variations $\delta_{g}\mathbf{\vartheta
}$ do not.

The set $\delta_{\mathrm{tr}}\mathbf{\vartheta}$ and $G_{\mathrm{tr}}$ is a
trivial symmetry. All the variations $\delta_{\mathrm{tr}}\mathbf{\vartheta}$
and the corresponding conserved charge $G_{\mathrm{tr}}$ vanish on the
extremals. The charge $G_{\mathrm{tr}}$ depends on the extremals as
$G_{\mathrm{tr}}=O\left(  I^{2}\right)  .$

In what follows, we present a constructive way for finding the components of
the decomposition (\ref{sdc.24}).

\subsubsection{Constructing the global canonical part of a symmetry}

Assuming that $\delta\mathbf{\vartheta}$ and $G$ is a symmetry, and taking the
structure of the total Hamiltonian in the case under consideration into
account, we can write the symmetry equation (\ref{I.12}) as%
\begin{equation}
\delta\vartheta E^{-1}\mathcal{I}-U^{\left(  1\right)  }\delta\lambda
_{U}-\lambda_{U}\delta U^{\left(  1\right)  }-\mathcal{P}^{(1)}\delta
\lambda_{\mathcal{P}}+\frac{dG}{dt}=0, \label{sdc.26}%
\end{equation}%
\[
I=(\Omega,J)=O\left(  \frac{\delta S_{\mathrm{H}}}{\delta\vartheta}\right)
,\;J=(\mathcal{I},\;\lambda_{U}),\;\mathcal{I}=\overset{\cdot}{\vartheta
}-\left\{  \vartheta,H\right\}  -\left\{  \vartheta,\mathcal{P}^{(1)}\right\}
\lambda_{\mathcal{P}}\,,
\]
We let denote via $\delta_{J}\mathbf{\vartheta}$, $G_{J}^{\prime}$ the
corresponding zero-order terms in the decomposition of the quantities
$\delta\mathbf{\vartheta}$, $G$ with respect to the extremals $J.$ We have%
\begin{equation}
\left(
\begin{array}
[c]{c}%
\delta\mathbf{\vartheta}\\
G
\end{array}
\right)  =\left(
\begin{array}
[c]{c}%
\delta_{J}\mathbf{\vartheta}(\eta,\lambda_{\mathcal{P}}^{[]})+O(J)\\
G_{J}^{\prime}(\eta,\lambda_{\mathcal{P}}^{[]})+B_{m}(\omega,\lambda
_{\mathcal{P}}^{[]})J^{[m]}+O(J^{2})
\end{array}
\right)  . \label{sdc.25}%
\end{equation}
We then rewrite the equation (\ref{sdc.26}) retaining only the terms of the
zero and first order with respect to the extremals $J$. We obtain%
\begin{align}
&  \delta_{J}\vartheta E^{-1}\mathcal{I}-\mathcal{P}^{(1)}\delta
\lambda_{\mathcal{P}}=-\hat{H}G_{J}^{\prime}+\left\{  \mathcal{P}^{(1)}%
,G_{J}^{\prime}\right\}  \lambda_{\mathcal{P}}+\lambda_{U}\{U^{\left(
1\right)  },G_{J}^{\prime}\}\nonumber\\
&  +\left\{  \vartheta,G_{J}^{\prime}\right\}  E^{-1}\mathcal{I}%
-J^{[m]}\hat{H}B_{m}+\lambda_{\mathcal{P}}\left\{  \mathcal{P}^{(1)}%
,B_{m}\right\}  J^{[m]}-B_{m}J^{[m+1]}+O(\Omega I)\,. \label{sdc.27}%
\end{align}
Here, the contributions from the terms $U^{\left(  1\right)  }\delta
\lambda_{U}$ and $\delta\lambda_{U}U^{\left(  1\right)  }$ are accumulated in
the term $O(\Omega I)$, and the operator $\hat{H}$ is defined by
\begin{equation}
\hat{H}F=\left\{  F,H\right\}  +\left(  \frac{\partial}{\partial t}%
+\lambda^{\left[  m+1\right]  }\frac{\partial}{\partial\lambda^{\left[
m\right]  }}\right)  F\,. \label{sdb.12}%
\end{equation}

Analyzing terms with the extremals $J^{[m]}$ (beginning with the highest
derivative) in Eq. (\ref{sdc.27}), we can see that all $B_{m}=0$. Considering
then terms proportional to $\mathcal{I}$ in Eq. (\ref{sdc.27}), we obtain the
expression
\[
\delta_{J}\vartheta=\left\{  \vartheta,G_{J}^{\prime}\right\}  +O(\Omega)
\]
for the variation $\delta_{J}\vartheta.$ Then, taking the relations%
\[
\delta\Omega=O(I)\Longrightarrow\delta_{J}\Omega=O(\Omega)=\left\{
\Omega,G_{J}^{\prime}\right\}  +O(\Omega)
\]
into account, we can see that%
\begin{equation}
\left\{  \Omega,G_{J}^{\prime}\right\}  =O(\Omega)\,. \label{sdc.28}%
\end{equation}
We can verify that $\left\{  \mathcal{P},G_{J}^{\prime}\right\}  $ is a
first-class function, which means that%
\begin{equation}
\left\{  \mathcal{P},G_{J}^{\prime}\right\}  =O(\mathcal{P})+O(\Omega^{2})\,.
\label{sdc.29}%
\end{equation}
Considering the remaining terms in Eq. (\ref{sdc.27}), we obtain the equation%
\begin{equation}
\mathcal{P}^{(1)}\delta_{J}\lambda_{\mathcal{P}}=\hat{H}G_{J}^{\prime}%
+\lambda_{\mathcal{P}}\left\{  \mathcal{P}^{(1)},G_{J}^{\prime}\right\}
+O(\Omega^{2})\,, \label{sdc.30}%
\end{equation}
which relates $\delta_{J}\lambda_{\mathcal{P}}$ and $G_{J}^{\prime}$.

This equation (\ref{sdc.30}) allows studying the function $G_{J}^{\prime}$ in
more detail. For this, we rewrite this equation (taking (\ref{sdb.12}) and
(\ref{sdc.29}) into account) as%
\[
\left\{  G_{J}^{\prime},H\right\}  +\left(  \frac{\partial}{\partial
t}+\lambda_{\mathcal{P}}^{\left[  m+1\right]  }\frac{\partial}{\partial
\lambda_{\mathcal{P}}^{\left[  m\right]  }}\right)  G_{J}^{\prime
}=O(\mathcal{P})+O(\Omega^{2})\,.
\]
Analyzing the terms which contain Lagrange multipliers $\lambda_{\chi}^{[m]}$
(beginning with the highest derivative) in this equation, we can see that
these multipliers can enter only the terms that vanish on the constraint
surface. For example, considering terms with the highest derivative
$\lambda_{\mathcal{P}}^{\left[  M+1\right]  }$ in the latter equation, we get%
\[
\frac{\partial G_{J}^{\prime}}{\partial\lambda_{\mathcal{P}}^{\left[
M\right]  }}=O(\mathcal{P})+O(\Omega^{2})\Longrightarrow G_{J}^{\prime}%
=G_{J}^{\prime}(\lambda_{\mathcal{P}},\cdots,\lambda_{\mathcal{P}}^{\left[
M-1\right]  })+O(\mathcal{P})+O(\Omega^{2}).
\]
In the same manner we finally obtain: $G_{J}^{\prime}=G_{J}(\vartheta
)+O(\mathcal{P})+O(\Omega^{2}).$ With Eqs. (\ref{sdc.28}) and (\ref{sdc.29})
taken into account, this implies:%

\[
\left\{  U,G_{J}\right\}  =O(\Omega)\,,\;\left\{  \mathcal{P},G_{J}\right\}
=O(\mathcal{P})+O(\Omega^{2})\,.
\]

Therefore, the above consideration allows representing a refined version of
the representation (\ref{sdc.25})%
\begin{equation}
\left(
\begin{array}
[c]{c}%
\delta\vartheta\\
\delta\lambda_{U}\\
\delta\lambda_{\mathcal{P}}\\
G
\end{array}
\right)  =\left(
\begin{array}
[c]{c}%
\left\{  \vartheta,G_{J}+B_{\mathcal{P}}\mathcal{P}\right\}  +O(I)\\
O(I)\\
\delta_{J}\lambda_{\mathcal{P}}(\vartheta,\lambda_{\mathcal{P}}^{[]})+O(J)\\
G_{J}+B_{\mathcal{P}}\mathcal{P}+O(I^{2})
\end{array}
\right)  . \label{sdc.37}%
\end{equation}
where $B_{\mathcal{P}}=B_{\mathcal{P}}(\vartheta,\lambda_{\mathcal{P}}^{[]}),$
and the function $G_{J}(\vartheta)$ obey the relations
\begin{equation}
\left\{  G_{J},U\right\}  =O(\Omega)\,,\;\left\{  G_{J},\mathcal{P}\right\}
=O(\mathcal{P})+O(\Omega^{2})\,,\;\left\{  G_{J},H\right\}  =O(\mathcal{P}%
)+O(\Omega^{2})\,. \label{sdc.31}%
\end{equation}

We select from the function $G_{J}$ a part $G_{I}$ that does not vanish on the
constraint surface,%
\begin{equation}
G_{J}(\vartheta)=g\left(  \omega\right)  +g_{1}\left(  \omega,Q\right)
Q+O\left(  \Omega\right)  \,. \label{sdc.33}%
\end{equation}
Because of the relation (\ref{sdc.31}), the function $g_{1}\left(
\omega,Q\right)  $ in (\ref{sdc.33}) is zero, and, moreover, $O\left(
\Omega\right)  =O\left(  \mathcal{P}\right)  +O\left(  \Omega^{2}\right)  .$
We define $G_{I}\left(  \vartheta\right)  $ as%
\begin{equation}
G_{I}\left(  \vartheta\right)  =g\left(  \omega\right)  \,. \label{sdc.34}%
\end{equation}
We then have%
\[
G_{J}(\vartheta)=G_{I}(\vartheta)+G_{1}(\vartheta),\;G_{1}(\vartheta
)=O(\mathcal{P})+O(\Omega^{2})\,.
\]
Therefore, in virtue of (\ref{sdc.31}),%
\begin{align}
&  G=G_{I}(\vartheta)+O(\mathcal{P})+O(I^{2})\,,\nonumber\\
&  \hat{H}G_{I}=0,\;\left\{  U,G_{I}\right\}  =\left\{  \mathcal{P}%
,G_{I}\right\}  =0\,. \label{sdc.35}%
\end{align}

We now define the variations $\delta_{I}\mathbf{\vartheta}$ as%
\begin{align}
&  \delta_{I}\vartheta=\left\{  \vartheta,G_{I}\right\}  \Longrightarrow
\delta_{I}\omega=\left\{  \omega,G_{I}\right\}  \,,\;\delta_{I}Q=\delta
_{I}\mathcal{P}=\delta_{I}U=0\,,\nonumber\\
&  \delta_{I}\lambda_{U}=\delta_{I}\lambda_{\mathcal{P}}=0\,. \label{sdc.36}%
\end{align}

The set $\delta_{I}\mathbf{\vartheta}$, $G_{I}$ is an exact symmetry of the
action $S_{\mathrm{H}}$. In what follows, this symmetry is denoted by%
\[
\delta_{I}\vartheta=\delta_{c}\vartheta=\{\vartheta,G_{c}\},\;\,\delta
_{I}\lambda=\delta_{c}\lambda=0\,,\;G_{I}=G_{c}=g(\omega)\,.
\]

\subsubsection{Constructing the gauge and the trivial parts of a symmetry}

At this step we represent a symmetry $\delta\mathbf{\vartheta}$, $G$ as%
\begin{equation}
\delta\mathbf{\vartheta}=\delta_{c}\mathbf{\vartheta}+\delta_{r}%
\mathbf{\vartheta},\;G=G_{c}+G_{r}\,. \label{sdc.41}%
\end{equation}
Because $\delta_{c}\mathbf{\vartheta}$, $G_{c}$ is a symmetry, it is obvious
that $\delta_{r}\mathbf{\vartheta}$, $G_{r}$ is also a symmetry. Using Eqs.
(\ref{sdc.31}), we can verify that the following relations%
\begin{align}
&  G_{r}=\sum_{i=1}^{\aleph_{\mathcal{P}}}\sum_{a=i}^{\aleph_{\mathcal{P}}%
}K_{i|a}\left(  \omega,Q,\lambda_{\mathcal{P}}^{[]}\right)  \mathcal{P}%
^{\left(  i|a\right)  }\,+O\left(  I^{2}\right)  \,,\nonumber\\
&  \delta_{r}\eta=\sum_{i=1}^{\aleph_{\mathcal{P}}}\sum_{a=i}^{\aleph
_{\mathcal{P}}}\left\{  \eta,\mathcal{P}^{\left(  i|a\right)  }\right\}
K_{i|a}\left(  \omega,Q,\lambda_{\mathcal{P}}^{[]}\right)  +O\left(  I\right)
\,, \label{sdc.42}%
\end{align}
where $K$ are some local functions, hold.

In turn, we represent the symmetry $\delta_{r}\mathbf{\vartheta}$, $G_{r}$ in
the following form%
\begin{equation}
\delta_{r}\mathbf{\vartheta}=\delta_{\bar{\nu}}\mathbf{\vartheta}%
+\delta_{\mathrm{tr}}\mathbf{\vartheta}\,,\;G_{r}=G_{\bar{\nu}}+G_{\mathrm{tr}%
}\,, \label{sdc.43}%
\end{equation}
where the set $\delta_{\bar{\nu}}\mathbf{\vartheta}$, $G_{\bar{\nu}}$ is the
gauge transformation given by Eqs. (\ref{sdc.4}), and (\ref{sdc.7}) with
specific fixed values of the gauge parameters,
\begin{equation}
\nu_{i}=\bar{\nu}_{i}\left(  t,\eta,\lambda^{\lbrack]}\right)  =K_{i|i}\left(
\omega,Q,\lambda_{\mathcal{P}}^{[]}\right)  ,\, \label{sdc.44}%
\end{equation}
that are either identically zero or do not vanish on the constraint surface.
This implies that%
\begin{equation}
G_{\bar{\nu}}=O\left(  \mathcal{P}\right)  +O\left(  I^{2}\right)
\,,\;\delta_{\bar{\nu}}\vartheta=\left\{  \vartheta,G_{\bar{\nu}}\right\}  \,.
\label{sdc.45}%
\end{equation}
We must emphasize that by construction, the functions $K_{i|i}$ (and therefore
the gauge transformations) are identically zero whenever they vanish on the
constraint surface.

It follows from Eqs. (\ref{sdc.42}) that $\delta_{\mathrm{tr}}%
\mathbf{\vartheta}$, $G_{\mathrm{tr}}$ is a symmetry whose charge is of the
form%
\[
G_{\mathrm{tr}}=G_{\mathrm{tr}}^{\prime}+O(I^{2})\,,\;G_{\mathrm{tr}}^{\prime
}=\sum_{i=1}^{\aleph_{\mathcal{P}}-1}\sum_{a=i+1}^{\aleph_{\mathcal{P}}%
}K_{i|a}\left(  \omega,Q,\lambda_{\mathcal{P}}^{[]}\right)  \mathcal{P}%
^{\left(  i|a\right)  }\,.
\]
In what follows, we will see that $\delta_{\mathrm{tr}}\mathbf{\vartheta}$,
$G_{\mathrm{tr}}$ is a trivial symmetry. For the symmetry $\delta
_{\mathrm{tr}}\mathbf{\vartheta}$, $G_{\mathrm{tr}}$ we write a decomposition
of the form (\ref{sdc.25}),%
\begin{equation}
\left(
\begin{array}
[c]{c}%
\delta\mathbf{\vartheta}_{\mathrm{tr}}\\
G_{\mathrm{tr}}%
\end{array}
\right)  =\left(
\begin{array}
[c]{c}%
\delta_{\mathrm{tr}J}\mathbf{\vartheta}(\omega,Q,\lambda_{\mathcal{P}}%
^{[]})+O(J)\\
G_{\mathrm{tr}J}^{\prime}(\vartheta,\lambda_{\mathcal{P}}^{[]})+O(J^{2}%
),\;G_{\mathrm{tr}J}^{\prime}=G_{\mathrm{tr}}^{\prime}+O(\Omega^{2})
\end{array}
\right)  \label{sdc.46}%
\end{equation}
taking into account that $B_{m}=O(\Omega).$ All the relations that hold for
the quantities $\delta_{J}\mathbf{\vartheta}$, $G_{J}$ also hold for the
quantities $\delta_{\mathrm{tr}J}\mathbf{\vartheta}$, $G_{\mathrm{tr}%
J}^{\prime}$. In particular, the charge $G_{\mathrm{tr}}^{\prime}$ obeys the
equation%
\begin{equation}
\mathcal{P}^{(1)}\delta_{\mathrm{tr}}^{\prime}\lambda_{\chi}%
=\hat{H}G_{\mathrm{tr}}^{\prime}+\lambda_{\mathcal{P}}\left\{  \mathcal{P}%
^{(1)},G_{\mathrm{tr}}^{\prime}\right\}  +O(\Omega^{2}),\;\delta_{\mathrm{tr}%
}^{\prime}\lambda_{\mathcal{P}}=\left.  \delta_{\mathrm{tr}}\lambda
_{\mathcal{P}}\right|  _{I=0}=\delta_{\mathrm{tr}J}\lambda_{\mathcal{P}%
}+O(\Omega), \label{sdc.47}%
\end{equation}
which is similar to Eq. (\ref{sdc.30}). The equation (\ref{sdc.47}) implies
the following equation for the local functions $K_{i|a}\,,\;a=i+1,...,\aleph
_{\mathcal{P}}$:%
\[
\sum_{i=1}^{\aleph_{\mathcal{P}}-1}\sum_{a=i+1}^{\aleph_{\mathcal{P}}}\left(
\mathcal{P}^{\left(  i|a\right)  }\hat{H}K_{i|a}+K_{i|a}\mathcal{P}^{\left(
i+1|a\right)  }\right)  =\mathcal{P}^{(1)}\delta_{\mathrm{tr}}^{\prime}%
\lambda_{\mathcal{P}}+O(\Omega^{2})\,.
\]
Considering this equation on the constraint surface $\Omega^{\left(
...\aleph_{\mathcal{P}}-1\right)  }=0$, we obtain that%

\[
K_{\aleph_{\mathcal{P}}-1|\aleph_{\mathcal{P}}}\mathcal{P}^{\left(
\aleph_{\mathcal{P}}|\aleph_{\mathcal{P}}\right)  }=O(\Omega^{2}%
)\Longrightarrow K_{\aleph_{\mathcal{P}}-1|\aleph_{\mathcal{P}}}=0\,.
\]
Substituting the expression for $K_{\aleph_{\mathcal{P}}-1|\aleph
_{\mathcal{P}}}$ into Eq. (\ref{sdc.47}), and considering the resulting
equation on the constraint surface $\Omega^{\left(  ...\aleph_{\mathcal{P}%
}-2\right)  }=0,$ we obtain $K_{\aleph_{\mathcal{P}}-2|\aleph_{\mathcal{P}}%
}=0$, and so on. We thus, see that all $K_{i|a}=0$, $a=i+1,...,\aleph
_{\mathcal{P}},$ and therefore%
\begin{equation}
G_{\mathrm{tr}}=O(I^{2})\,. \label{sdc.48}%
\end{equation}
It then follows from Eq. (\ref{sdc.47})%
\[
\mathcal{P}^{(1)}\delta_{\mathrm{tr}}^{\prime}\lambda_{\mathcal{P}}%
=O(\Omega^{2})\Longrightarrow\delta_{\mathrm{tr}}^{\prime}\lambda
_{\mathcal{P}}=O(\Omega)\Longrightarrow\delta_{\mathrm{tr}}\lambda
_{\mathcal{P}}=O(I)\,.
\]

By construction, the transformation $\delta_{\mathrm{tr}J}$ is similar to
$\delta_{J}$. Therefore, the relation (\ref{sdc.31}) holds true for this
transformation and implies that%
\[
\delta_{\mathrm{tr}J}\vartheta=\{\vartheta,G_{\mathrm{tr}}^{\prime}\}+O\left(
\Omega\right)  =O(\Omega)\,,\;\;\delta_{\mathrm{tr}J}\lambda_{U}=O(\Omega)\,.
\]
Therefore,%
\begin{equation}
\delta_{\mathrm{tr}}\mathbf{\vartheta}=O(I)\,. \label{sdc.49}%
\end{equation}
The relations (\ref{sdc.48}) and (\ref{sdc.49}) prove that the symmetry
$\delta_{\mathrm{tr}}\mathbf{\vartheta}$, $G_{\mathrm{tr}}$ is trivial.

\section{Physical functions}

First of all, we recall the general understanding that physics can be
described in terms of gauge theories \cite{GitTy90}. Let the time evolution of
a classical system be given by genuine trajectories $\kappa\left(  t\right)  $
in the configuration space. The latter are solutions of the equations of
motion of the theory. On the other hand, the state of the classical system at
any given time instant $t$ is characterized by the set $\kappa^{\lbrack
]}\left(  t\right)  =\left(  \kappa^{\left[  l\right]  }\left(  t\right)
,\;l\geq0\right)  ,$ at this time instant, i.e., by a point in the jet space.
The trajectory in the configuration space creates a trajectory in the jet
space. The latter trajectory can be called the trajectory of system states. We
call two trajectories in the configuration space intersecting if the
corresponding trajectories in the jet space intersect at a given time instant.
Using such a terminology and the results of the Sect. III, we can say that
intersecting trajectories do exist in gauge theories . On the other hand, we
believe that for classical systems, we can introduce the notion of the system
physical state at each time instant, such that there exists a causal evolution
of the physical states in time. Namely, once a physical state is given at a
certain time, at all other times the physical states are determined in a
unique way. All the physical quantities are single-valued functions of the
physical state at a given time instant. The physical state is completely
determined as soon as all possible physical quantities are given in a certain
time instant. Therefore, at a first glance, there is a disagreement between
the causal evolution of the physical states and the absence of the causal
evolution of trajectories in the jet space for gauge theories. To eliminate
this discrepancy and to be able describe classical systems consistently with
the use of gauge theories, we can resort to the following natural interpretation:

a) Physical states of a classical system and, therefore, all local physical
quantities are uniquely determined by points of genuine trajectories in the
jet space.

b) All the functions that are used to describe physical quantities must
coincide at equal-time points of intersecting genuine trajectories in the jet space.

Item (b) ensures independence of the physical quantities from the
arbitrariness inherent to solutions of a gauge theory and reconciles item (a)
with the causal development of the physical states in time. Item (b) imposes
limitations on the possible form of these functions. The local functions that
obey item (b) are called physical functions. Suppose the local functions
$\mathcal{A}_{\text{\textrm{ph}}}\left(  \kappa^{\lbrack]}\right)  $ are
physical. This implies that for two arbitrary genuine intersecting
trajectories $\kappa$ and $\kappa^{\prime}$ the equality
\begin{equation}
\mathcal{A}_{\text{\textrm{ph}}}\left(  \kappa^{\lbrack]}\right)
=\mathcal{A}_{\text{\textrm{ph}}}\left(  \kappa^{\prime\lbrack]}\right)
\label{pia.5}%
\end{equation}
holds at any time instant.

We consider local physical functions in the Hamiltonian formulation and in the
special phase-space variables $\mathbf{\vartheta}$. Taking the equations of
motion (\ref{c.1}), (\ref{c.2}), and $\Omega=0$ into account, we can conclude
that any physical local functions of the form $\mathcal{A}_{\text{\textrm{ph}%
}}\left(  \mathbf{\vartheta}^{\left[  {}\right]  }\right)  $ can be
represented as%
\begin{equation}
\mathcal{A}_{\text{\textrm{ph}}}\left(  \mathbf{\vartheta}^{[]}\right)
=\emph{a}_{\mathrm{ph}}\left(  \omega,Q,\lambda_{\mathcal{P}}^{[]}\right)
+O\left(  \frac{\delta S}{\delta\mathbf{\vartheta}}\right)  \,. \label{pia.7}%
\end{equation}
It is now easy to establish restrictions on the functions $\emph{a}%
_{\mathrm{ph}}$ that follow from the condition (\ref{pia.5}) of physicality.
For this, we recall that there exist two intersecting at $t=0$ genuine
trajectories $\mathbf{\vartheta}$ and $\mathbf{\vartheta}^{\prime}$ such that
at the time instant $t$ they, having the same $\omega,$ and differ only by the
values of the variables $Q$ and $\lambda_{\mathcal{P}}^{\left[  l\right]  }$.
Namely,%
\begin{align}
&  \mathbf{\vartheta}\left(  t\right)  \Longrightarrow\left(  Q,\;\lambda
_{\mathcal{P}}^{[]}\right)  \,,\nonumber\\
&  \mathbf{\vartheta}^{\prime}\left(  t\right)  \Longrightarrow\left(
Q+\delta Q,\;\lambda_{\mathcal{P}}^{[]}+\delta\lambda_{\mathcal{P}}%
^{[]}\right)  \,, \label{pia.8}%
\end{align}
where all the quantities $Q,\;\lambda_{\mathcal{P}}^{[]},\delta Q,$ and
$\delta\lambda_{\mathcal{P}}^{[]}\;$are arbitrary. The existence of such
intersecting trajectories follows from the above consideration. The relation
(\ref{pia.5}) for such two intersecting trajectories implies the relation
\begin{equation}
\emph{a}_{\mathrm{ph}}\left(  \omega\left(  t\right)  ,Q,\;\lambda
_{\mathcal{P}}^{[]}\right)  =\emph{a}_{\mathrm{ph}}\left(  \omega\left(
t\right)  ,Q+\delta Q,\;\lambda_{\mathcal{P}}^{[]}+\delta\lambda_{\mathcal{P}%
}^{[]}\right)  \,. \label{pia.9}%
\end{equation}
for the function $\emph{a}_{\mathrm{ph}}.$ Because of the arbitrariness of the
quantities $Q,\;\lambda_{\mathcal{P}}^{[]},\delta Q,$ and $\delta
\lambda_{\mathcal{P}}^{[]},$ we obtain from the equation (\ref{pia.9}) that%
\begin{equation}
\frac{\partial\emph{a}_{\mathrm{ph}}}{\partial Q}=\frac{\partial
\emph{a}_{\mathrm{ph}}}{\partial\lambda^{\lbrack]}}=0\Longrightarrow
\emph{a}_{\mathrm{ph}}=\emph{a}_{\mathrm{ph}}\left(  \omega\right)  \,.
\label{pia.10}%
\end{equation}
Therefore, physical local functions of the form $\mathcal{A}%
_{\text{\textrm{ph}}}\left(  \mathbf{\vartheta}^{[]}\right)  $ can be
represented as%
\begin{equation}
\mathcal{A}_{\text{\textrm{ph}}}\left(  \mathbf{\vartheta}^{[]}\right)
=\emph{a}_{\mathrm{ph}}\left(  \omega\right)  +O\left(  \frac{\delta
S_{\mathrm{H}}}{\delta\mathbf{\vartheta}}\right)  \,. \label{pia.11}%
\end{equation}
In terms of the initial phase-space variables $\mathbf{\eta=}\left(
\eta,\lambda\right)  ,\;\eta=\left(  q,p\right)  $, any physical local
functions of the form $A_{\text{\textrm{ph}}}\left(  \mathbf{\eta}%
^{[]}\right)  $ has the structure%
\begin{equation}
A_{\text{\textrm{ph}}}\left(  \mathbf{\eta}^{[]}\right)  =a_{\mathrm{ph}%
}\left(  \eta\right)  +O\left(  \frac{\delta S_{\mathrm{H}}}{\delta
\mathbf{\eta}}\right)  \,. \label{pia.12}%
\end{equation}
It follows from (\ref{pia.11}) that bearing in mind that the set of
constraints $\mathcal{P}$ is equivalent to all FCC $\chi\left(  \eta\right)  $
in the initial phase-space variables, and that the set of constraints $\Omega$
is equivalent to all the initial constraints $\Phi\left(  \eta\right)  ,$ one
can write the physicality conditions for the functions $a_{\text{\textrm{ph}}%
}\left(  \eta\right)  $:%
\begin{equation}
\frac{\partial a_{\mathrm{ph}}}{\partial Q}=O\left(  \Omega\right)
\Longleftrightarrow\{a_{\mathrm{ph}},\chi\}=O\left(  \Phi\right)  \,.
\label{pia.13}%
\end{equation}
We are going to call conditions (\ref{pia.12}) and (\ref{pia.13}) the
physicality condition in the Hamiltonian sense. It is precisely in this sense
one has to understand the usual assertion that physical functions must commute
with first-class constraints on extremals. In fact, these conditions of
physicality are those which are usually called the Dirac conjecture.

On the other hand, in the Lagrangian formulation it is known that physical
functions must be gauge invariant on the extremals, see e.g. \cite{GitTy90}.
Such a condition is called the physicality condition in the Lagrangian sense.
Below, we are going to demonstrate the equivalence of these two conditions.

Let local functions $A=A\left(  \mathbf{\eta}^{[]}\right)  $ be physical in
the Hamiltonian sense. Consider their gauge variation $\delta\mathcal{A}$.
Such a variation has the following form, having (\ref{pia.12}) in mind,%
\begin{equation}
\delta A=\delta a\left(  \eta\right)  +O\left(  \frac{\delta S_{\mathrm{H}}%
}{\delta\mathbf{\eta}}\right)  \,. \label{pia.15}%
\end{equation}
Here we have used the fact that gauge variations of extremals are proportional
to extremals. Let us consider $\delta a$ taking into account (\ref{sdc.7}) and
(\ref{sdc.4}). Then one easily sees that%
\[
\delta a=\{a,G\}=O\left(  \left\{  a,\chi\right\}  \right)  +O\left(
\frac{\delta S_{\mathrm{H}}}{\delta\mathbf{\eta}}\right)  \,.
\]
Taking into account (\ref{pia.13}), we obtain that gauge variations of
physical functions are proportional to extremals,%
\begin{equation}
\delta A=O\left(  \frac{\delta S_{\mathrm{H}}}{\delta\mathbf{\eta}}\right)
\,. \label{pia.16}%
\end{equation}

Let now the local functions $A=A\left(  \mathbf{\eta}^{[]}\right)  $ be
physical in the Lagrangian sense, i.e. they obey Eq. (\ref{pia.16}). One can
always represent them in the form%
\[
A=f(\eta,\lambda_{\mathcal{P}}^{[]})+O\left(  \frac{\delta S_{\mathrm{H}}%
}{\delta\mathbf{\eta}}\right)  .
\]
The condition (\ref{pia.16}) implies:%
\begin{equation}
\{f,G\}+\sum_{m=0}^{m_{\max}}\frac{\partial f}{\partial\lambda_{\mathcal{P}%
}^{[m]}}\delta\lambda_{\mathcal{P}}^{[m]}=O\left(  \frac{\delta S_{\mathrm{H}%
}}{\delta\mathbf{\eta}}\right)  . \label{sa}%
\end{equation}
Let us consider those terms containing the highest time-derivatives of the
gauge parameters in the left-hand side of (\ref{sa}). Taking into account that
$\delta\lambda_{\mathcal{P}}^{a}=\nu_{a}^{[a]},$ see (\ref{sdc.211}), and the
fact that $G$ contains only the time derivatives $\nu_{a}^{[l]}\,,\;l<a$),
such terms have the form:%
\[
\sum_{a}^{\aleph_{\chi}}\frac{\partial f}{\partial\lambda_{\mathcal{P}%
}^{a[m_{\max}]}}\nu_{a}^{[a+m_{\max}]}\,.
\]
These terms have to be proportional to the extremals, which implies%
\[
\frac{\partial f}{\partial\lambda_{\mathcal{P}}^{a[m_{\max}]}}=O\left(
\frac{\delta S_{\mathrm{H}}}{\delta\mathbf{\eta}}\right)  \,.
\]
Similarly, we can verify that the function $f$ does not contain any $\lambda$
on the extremals, i.e.,%
\[
f(\eta,\lambda_{\chi}^{[]})=a\left(  \eta\right)  +O\left(  \frac{\delta
S_{\mathrm{H}}}{\delta\mathbf{\eta}}\right)  \,.
\]
Therefore,%
\begin{equation}
\mathcal{A}=a(\eta)+O\left(  \frac{\delta S_{\mathrm{H}}}{\delta\mathbf{\eta}%
}\right)  \,. \label{sdc.22}%
\end{equation}
Considering the equation (\ref{pia.16}) for the function (\ref{sdc.22}), we
obtain that%

\[
\{a,G\}=\sum_{i=1}^{\aleph_{\chi}}\sum_{b=i}^{\aleph_{\chi}}\left\{
a,\mathcal{P}^{(i|b)}\right\}  \nu_{b}^{[b-i]}=O\left(  \frac{\delta
S_{\mathrm{H}}}{\delta\mathbf{\eta}}\right)  \,,
\]
which implies
\[
\left\{  a,\mathcal{P}^{(i|b)}\right\}  =O\left(  \frac{\delta S_{\mathrm{H}}%
}{\delta\mathbf{\eta}}\right)  =O(\Phi)\,.
\]
because of the independence between $\nu_{b}^{[b-i]}$. This completes the
proof of the equivalence of the two definitions of physical functions.

\section{Conclusion}

We summarize the main conclusions.

Any continuous symmetry transformation can be represented as a sum of three
kind of symmetries, a global symmetry, a gauge symmetry, and a trivial
symmetry. If the global part of a symmetry and the corresponding canonical
charge are not identically zero, they do not vanish on the extremals. The
determination of the canonical charge from the corresponding equation, and
therefore the determination of the canonical part of a symmetry transformation
is ambiguous. However, we must understand that the ambiguity in the canonical
part of a symmetry transformation is always a sum of a gauge transformation
and a trivial transformation. The gauge part of a symmetry does not vanish on
the extremals, but the gauge charge vanishes on the extremals. We emphasize
that the gauge charge necessarily contains a part that is linear in the FCC,
and the remaining part of the gauge charge is quadratic in the extremals. The
trivial part of any symmetry vanishes on the extremals and the corresponding
charge is quadratic in the extremals.

The reduction of global symmetry transformations to the extremals are global
canonical symmetries of the physical action whose conserved charge is the
reduction of the complete conserved charge to the extremals.

Any global symmetry of the physical action is a global symmetry of the
complete Hamiltonian action.

The gauge transformations, taken on the extremals, only transform the
nonphysical variables $Q$ and $\lambda_{\mathcal{P}}\,.$

We can see that there are no gauge transformations which cannot be represented
in the form (\ref{sdc.4}). This follows from the structure of arbitrary
symmetry transformation presented above. Namely, as was demonstrated, any
symmetry transformation whose charge vanishes on the extremals is a sum of a
particular gauge transformation and of a trivial transformation.

The gauge charge contains time derivatives of the gauge parameters whenever
there exist secondary FCC.

Another assertion holds. We can see that the numbers of nonphysical variables
both in Lagrangian and Hamiltonian formulations are respectively equal to the
complete numbers of gauge parameters and their time derivatives that enter in
the gauge transformations in these formulations. Indeed, in the Lagrangian
formulation, the number of the nonphysical coordinates coincide with the
number of the FCC in Hamiltonian formulation (with the number of the variables
$Q$), and, therefore, coincide with the complete number of gauge parameters
and their time derivatives that enter the gauge transformations of the
coordinates in the Lagrangian formulation. In the Hamiltonian formulation, the
nonphysical variables are both $Q$ and $\lambda_{\chi}.$ At the same time, in
this formulation, the gauge transformations of the Lagrange multipliers
$\lambda_{\chi}$ contain an additional time derivative in comparison with the
gauge transformations of the coordinates in the Lagrangian formulation. The
number of $\lambda_{\chi}$ is equal to the number of primary FCC and,
therefore, is equal to the number of the gauge parameters. A simple estimation
confirms the above assertion.

In the same manner as in Sect. II, we can demonstrate that there is a choice
of superspecial phase-space variables that already includes the variables $U$
and which significantly simplifies the Hamiltonian $H_{\mathrm{SCC}}^{(1)}.$
Namely, for such a choice the variables $U$ have the structure $U=(V;\,u),$
where both $V$ and$\,u$ are sets of pairs of conjugate coordinates and
momenta. The variables from these sets are divided into groups according to
the stages of the Dirac procedure and organized in chains (labeled by the
index $a$). The variables $V$ consist of coordinates $\Theta$ and conjugate
momenta $\Pi,$ namely,
\begin{align*}
V  &  =(\Theta_{\mu_{a}}^{(i|2a)},\,\Theta_{\nu_{a},s}^{(i|2a+1)};\Pi_{\mu
_{a}}^{(i|2a)},\,\Pi_{\nu_{a},s}^{(i|2a+1)})\,,\;\;1\leq a\leq\aleph_{\varphi
}/2\,,\;i=1,...,a\,,\;s=1,2,\\
u  &  =\left(  u_{\zeta,s}^{(1)}\,,\;u_{\nu_{a},s}^{(2a+1)}\right)  .
\end{align*}
The variables $u^{(1)}$ are primary constraints (first-stage constraints); the
variables $u^{(2a+1)}$ are $2a+1-$stage constraints; the variables
$\Pi^{(i|2a)}$ and $\,\Pi^{(i|2a+1)}$ are $i$-stage constraints; the variables
$\Theta^{(i|2a)}$ are $2a-(i-1)$-stage constraints; the variables
$\Theta^{(i|2a+1)}$ are $2a+1-(i-1)$-stage constraints.

The variables are divided in even and odd chains$.$ Variables in even chains
(labeled by $2a$) are labeled by the index $\mu_{a}$ , variables in odd chains
(labeled by $1$, $2a+1$) are labeled by the index $\zeta$, $\nu_{a}$ and by
the index $s.$ The number of the indices $\mu_{a}$ and $\zeta$, $\nu_{a}$ can
be equal to zero.

In terms of the variables $V,u$ the Hamiltonian $H_{\mathrm{SCC}}^{(1)}$
becomes:%
\begin{align*}
&  H_{\mathrm{scc}}^{(1)}=h_{\mathrm{odd}}+h_{\mathrm{even}}+\lambda
^{(1)}u^{(1)}\,,\\
&  h_{\mathrm{even}}=\sum_{a=1}\left(  \sum_{i=1}^{a-1}\Theta^{(i|2a)}%
\Pi^{(i+1|2a)}+\sigma_{2a}(\Theta^{(i|2a)})^{2}+\lambda^{(2a)}\Pi
^{(1|2a)}\right)  \,,\\
&  h_{\mathrm{odd}}=\sum_{a=1}\left(  \sum_{i=1}^{a-1}\Theta^{(i|2a+1)}%
\Pi^{(i+1|2a+1)}+\sigma_{2a+1}\Theta^{(a|2a+1)}u^{(2a+1)}+\lambda^{(2a+1)}%
\Pi^{(1|2a+1)}\right)  \,,
\end{align*}
where $\sigma\neq0$ are some numbers. There is a summation over the indices
$\mu,\nu,$ and $\zeta,$ in particular $\sigma_{2a}(\Theta^{(i|2a)})^{2}%
=\sum_{\mu_{a}}\sigma_{2a,\mu_{a}}(\Theta_{\mu_{a}}^{(i|2a)})^{2}.$

In the refined superspecial phase-space variables, the consistency conditions
that start with the primary SCC require that all the corresponding Lagrange
multipliers $\lambda^{(1)},\lambda^{(2a)},$ and $\lambda^{(2a+1)}$ be zero.
See the scheme of constraint chains below%
\[%
\begin{array}
[c]{ccccccccccccccc}%
u_{s}^{(1)} & \rightarrow & \lambda_{s}^{(1)} &  &  &  &  &  &  &  &  &  &  &
& \\
\Pi^{(1|2)} & \rightarrow & \Theta^{(1|2)} & \rightarrow & \lambda^{(2)} &  &
&  &  &  &  &  &  &  & \\
\Pi_{s}^{(1|3)} & \rightarrow & \Pi_{s}^{(2|3)} & \rightarrow & u_{s}^{(3)} &
\rightarrow & \Theta_{s}^{(2|3)} & \rightarrow & \Theta_{s}^{(1|3)} &
\rightarrow & \lambda_{s}^{(3)} &  &  &  & \\
\Pi^{(1|4)} & \rightarrow & \Pi^{(2|4)} & \rightarrow & \Theta^{(2|4)} &
\rightarrow & \Theta^{(1|4)} & \rightarrow & \lambda^{(4)} &  &  &  &  &  & \\
\vdots &  &  &  &  &  &  &  &  &  &  &  &  &  & \\
\Pi^{(1|2a)} & \rightarrow & ... & \rightarrow & \Pi^{(a|2a)} & \rightarrow &
\Theta^{(a|2a)} & \rightarrow & ... & \rightarrow & \Theta^{(1|2a)} &
\rightarrow & \lambda^{(2a)} &  & \\
\Pi_{s}^{(1|2a+1)} & \rightarrow & ... & \rightarrow & \Pi_{s}^{(a|2a+1)} &
\rightarrow & u_{s}^{(2a+1)} & \rightarrow & \Theta_{s}^{(a|2a+1)} &
\rightarrow & ... & \rightarrow & \Theta_{s}^{(1|2a+1)} & \rightarrow &
\lambda_{s}^{(2a+1)}\\
\vdots &  &  &  &  &  &  &  &  &  &  &  &  &  &
\end{array}
\]

Finally, we must mention that we have demonstrated (using some natural
assumptions) the equivalence of two definitions of physicality. One of them
pertaining to the Lagrangian formulation, which states that physical functions
are gauge invariant on the extremals, and the other definition pertaining to
the Hamiltonian formulation, which requires that physical functions commute
with FCC (Dirac conjecture).

\begin{acknowledgement}
Gitman is grateful to the foundations FAPESP, CNPq for support and to the
Lebedev Physics Institute (Moscow) for hospitality; Tyutin thanks RFBR
02-02-16944 and LSS-1578-2003.2 for partial support.
\end{acknowledgement}

\end{document}